\documentclass[]{mn2e}
\usepackage{amsmath}
\usepackage{amssymb}
\usepackage{graphicx}

\def\kms{km~s$^{-1}$ }
\def\etal{et al.\,}

\def\eg{e.g., \,}

\title[The Double Galaxy Cluster Abell 2465 I.]
{The Double Galaxy Cluster 
Abell 2465 I. Basic Properties: Optical Imaging and Spectroscopy}

\author[G. A. Wegner]
{Gary A. Wegner$^1$\\
$^1$Department of Physics \& Astronomy, Dartmouth College, 6127 Wilder
Laboratory, Hanover, NH 03755, U.S.A.}

\date{Accepted ; in original form }

\begin{document}
%\label{firstpage}
\maketitle

\begin{abstract}
Optical imaging and spectroscopic observations of the $z = 0.245$
double galaxy cluster Abell 2465 are described. This object 
appears to be undergoing a major merger. It is a double X-ray
source and is detected in the radio at 1.4 GHz.
The purpose of this paper is to investigate signatures of the 
interaction of the two components. Redshifts were measured to
determine velocity dispersions and virial radii of each component.
The technique of fuzzy clustering was used to assign membership weights
to the galaxies in each clump. Using redshifts of 93 cluster members
within 1.4 Mpc of the subcluster centres, the virial mass of the NE
component is $M_{V} = 4.1 \pm 0.8 \times 10^{14} M_\odot$ and
$M_{V} = 3.8 \pm 0.8 \times 10^{14} M_\odot$ for the SW. These agree
within the errors with masses from X-ray scaling relations. 
The projected velocity
difference between the two subclusters is 205 $\pm$ 149 \kms.
The anisotropy parameter,
$\beta$, is found to be low for both components. Spectra of
37\% of the spectroscopically observed galaxies show
emission lines and are predominantly star forming in
the diagnostic diagram. No strong AGN sources were found.
The emission line
galaxies tend to lie between the two cluster centres
with more near the SW clump. The luminosity functions of the two subclusters 
differ. The NE component is similar to many rich clusters, while the SW 
component has more faint galaxies. 
The NE clump's light profile follows a single NFW profile with c = 10 
while the SW is
better fit with an extended outer region and a compact inner core,
consistent with available X-ray data indicating that the SW clump has
a cooling core. The observed differences and properties of the two components
of Abell 2465 are interpreted to have been caused by a collision
2-4 Gyr ago, after which they have moved apart and are now
near their apocentres, although the start of a merger remains a
possibility. The number of emission line galaxies gives weight to 
the idea that galaxy cluster collisions trigger star formation. 

\end{abstract}


\section{INTRODUCTION}

Fundamental questions of current 
astrophysics involve the roles of dark matter, baryonic matter, and
dark energy as driven by gravity in the formation of
the large-scale structure and galaxies. 
Double or multiple galaxy clusters can
potentially provide information on the dynamics and structure formation on
($r \ga$ 1 Mpc) scales and interest in them has grown from
both the standpoints of modelling and observation. 

Although the presence of substructure in galaxy clusters has long been known,
compared to single galaxy clusters, the properties of double and multiple
clusters have received less attention owing to their added complexity.
Interest in the observed 
substructure of galaxy clusters was pioneered by, \eg
Geller \& Beers (1982) and studies employing the radial infall 
model (Beers, Geller \& Huchra 1982; Beers \etal 1991) was used for rough
dynamical estimates. 

With the realization of their importance, a growing
number of systems have now been more fully studied dynamically, 
from weak lensing, and in X-rays. A partial list
includes: Abell 168 (Hallman \& Markevitch 2004), Abell 399/401 (Sakelliou \&
Ponman 2004), Abell 520 (Girardi \etal 2008), Abell 521 (Ferrari \etal 2003),
Cl0024+17 (Jee \etal 2007), the ``bullet cluster,'' 
1E0657-56 (Clowe et al. 2006), Abell 2146 (Russell \etal 2010), 
RXJ1347.5-1145 (Brada\v{c} \etal 2008a),
A399 and A401 (Yuan \etal 2005), Abell 2163 (Maurogordato \etal 2008),
Abell 85 (Tanaka \etal 2010), and Abell 901/902 (Heiderman \etal 2009).
Okabe \& Umetsu (2008) studied seven merging clusters using weak lensing.

Radio emission from merging clusters in the form of diffuse non-thermal 
radio halos or relics that arise from merger shocks in the interactions of 
the colliding galaxy clusters.has also been described by several groups
including Slee, Roy, \& Murgia (2001), Feretti (2002) who describe several 
objects, Bagchi \etal (2006), Abell 3376, Orr\'{u} \etal (2007), Abell 2744 
and Abell 2219, Bonafede \etal (2009), Abell 2345, van Weeren \etal (2009),
A2256. Skillman \etal (2010) summarize the the modeling situation.    

Modelling galaxy cluster mergers and collisions predict observable
signatures (\eg Roettiger \etal (1996, 1997; Ricker 1998; Tazikawa 2000;
Ricker \& Sarazin 2001; Ritchie \& Thomas 2002; 
Springel \& Farrar 2007; Mastropietro
\& Burkert 2008; Poole \etal 2008; Planelles \& Quilis 2009). 
These simulations have mostly
focused on the behavior of the baryonic and dark matter components and
use a range of initial profiles and conditions and impact parameters 
which include both
off-centre and head-on collsions. These calculations predict
differing behaviors for the baryonic and
dark matter components of the clusters at subsequent phases of the collisions.
In a typical merger, the dark matter and the baryonic gas are elongated
along the collision axis with a displacement between the baryonic and
dark matter components. The gas, in addition, is shocked which results in
multiple X-ray peaks and gas splashed perpendicularly to the direction of
the merger. This produces non-isothermal temperature distributions and
the increased ram pressure from the shocks could induce star formation in
the member galaxies as well as 'sloshing' (Markevitch \& Vikhlinin 2007).

Several authors have attempted to extract information from double and
multiple galaxy clusters on
the nature of gravity and dark matter on galactic cluster ($\sim$ 1 Mpc)
distance scales and
up to now this has mainly centred on analyzing the 1E0657-56
cluster. Farrar \& Rosen (2006), Brownstein \& Moffat (2007), Angus \&
McGaugh (2008), Schmidt, Vikhlinin, \& Hu (2009), and De Lorenci, 
Faundez-Abans, \& Pereira (2009) are among those who investigated whether
modifications to gravity are needed to fit the available dynamical data.
Springel \& Farrar (2007), Pointecouteau \& Silk (2005), 
and Hayashi \&White (2006), however indicated that
modifications are unneccesary. For studying the properties of 
dark matter, the situation is somewhat more definite. Clowe \etal (2006)
used weak lensing measurements of the bullet cluster to indicate direct
proof of the the presence of dark matter from the offset between the X-ray
gas and the lensing centres. Shan \etal (2010) have studied further offsets
between dark and ordinary matter in a further 38 lensed galaxy clusters. 
Galaxy clusters have been employed to place limits 
on neutrino masses (\eg Tremaine \& Gunn 1979; Angus, Famaey, \& Diaferio
2010; Natarajan \& Zhao 2008) and discussed whether or not
such particle masses are needed to save the modified Newtonian dynamics
formula.   

Even if one dismisses such claims, a considerable amount of 
more conventional information is obtainable
from double galaxy clusters. This includes 
possible modifications to luminosity functions, mass
profiles and velocity dispersion anisotropy measures (the $\beta$ parameter)
as a result of their interactions. Luminosity functions
contain information on the galaxy formation history
(\eg Bingelli, Sandage \& Tammann 1988) and 
have been studied in detail at a range
of redshifts and environments, mostly for single systems 
(\eg Wilson \etal 1997; De Propis \etal 2004; 
Blanton \etal 2003; Christlein \& Zabludoff 2003; Goto \etal 2005). Generally
single, and double Schechter functions (Schechter 1976), and Gaussian
functionshave been used to fit the LFs. Collisions may modify 
these properties compared to isolated single clusters at some level, but
this question about the effects of merging in double galaxy clusters,
whether or not their interactions produce
or lower star formation along with AGN activity,
has not been answered yet. Hwang \& Lee (2009) 
have reviewed empirical and theoretical evidence for this and conclude that
observations support the importance of mergers. 
Haines et al. (2009) and Chung (2010) have reported evidence
of enhanced star formation rates in interacting clusters including the 
bullet cluster. 

Mergers can distort galaxy cluster mass profiles.
Many investigators have compared theoretical mass profiles with
observations (\eg Biviano \& Girardi 2003; Katgert, Biviano, \& Mazure 2004;
Pointecouteau, Arnaud, \& Pratt 2005; Kubo \etal 2007; Okabe \& Umetsu 2008).
Although not all details of these models are agreed upon, the NFW
profile (Navarro, Frenk, \& White 1997) fits most observed profiles
within the virial radius with a
concentration parameter, $c$ for galaxy clusters is in
the range of $c = 4-6$ in agreement with theoretical results (\eg Zhao \etal
2003). In addition, for a spherical system with 
the NFW profile, the anisotropy parameter 
$\beta = 1-\sigma_{\theta}/\sigma_{r}$ (where $\sigma_{\theta}$ is the
azimuthal velocity dispersion and $\sigma_{r}$ the radial velocity
dispersion), is predicted to be near 0 at the
centre and to increase to about 0.3 beyond the virial radius and can provide
information on the properties of the dark matter (Host 2009). 

Many of the systems described in the literature are multiple and complex or
minor mergers where the mass of one component is considerably larger than the
other.The Abell 2465 double cluster discussed in this paper has a
relatively uncomplicated substructure and shows some 
evidence for either a past 
collision or a commencing merger between the two components. 
The mean redshift is $z = 0.245$.
ROSAT (Perlman et al. 2002), XMM (2008) data, and redshifts
show two physically related X-ray sources
5.5$\arcmin$ (1.2 Mpc) apart (hereafter the NE and SW clumps). As well, it
is a 1.4 GHz radio source in the NVSS (Condon \etal 1998). Both the
virial and X-ray masses obtained in this paper indicate that the mass ratio
of the two clumps is close to 1:1. Therefore the Abell 2465 is an example of 
a relatively rare major merger.  

This paper surveys the optical properties of 
Abell 2465 cluster and is organized as follows: Section 2 describes the
new imaging and spectroscopy, 
Section 3 gives estimates from spectroscopy of
virial masses and radii, values of $\beta$, and discusses available
X-ray and radio data and emission 
line galaxies in the two subclusters found from the spectra.
Section 4 describes the determination of
the luminosity functions, and Section 5 compares the
estimates of the light profile and the corresponding mass profile. Section
6 discusses the results in relation to a collision, 
and Section 7 lists the conclusions. 
The WMAP 5 year cosmological parameters are used
throughout this paper.

%\newpage
\begin{figure*}
\begin{center}
\leavevmode
\includegraphics[width=0.83\textwidth]{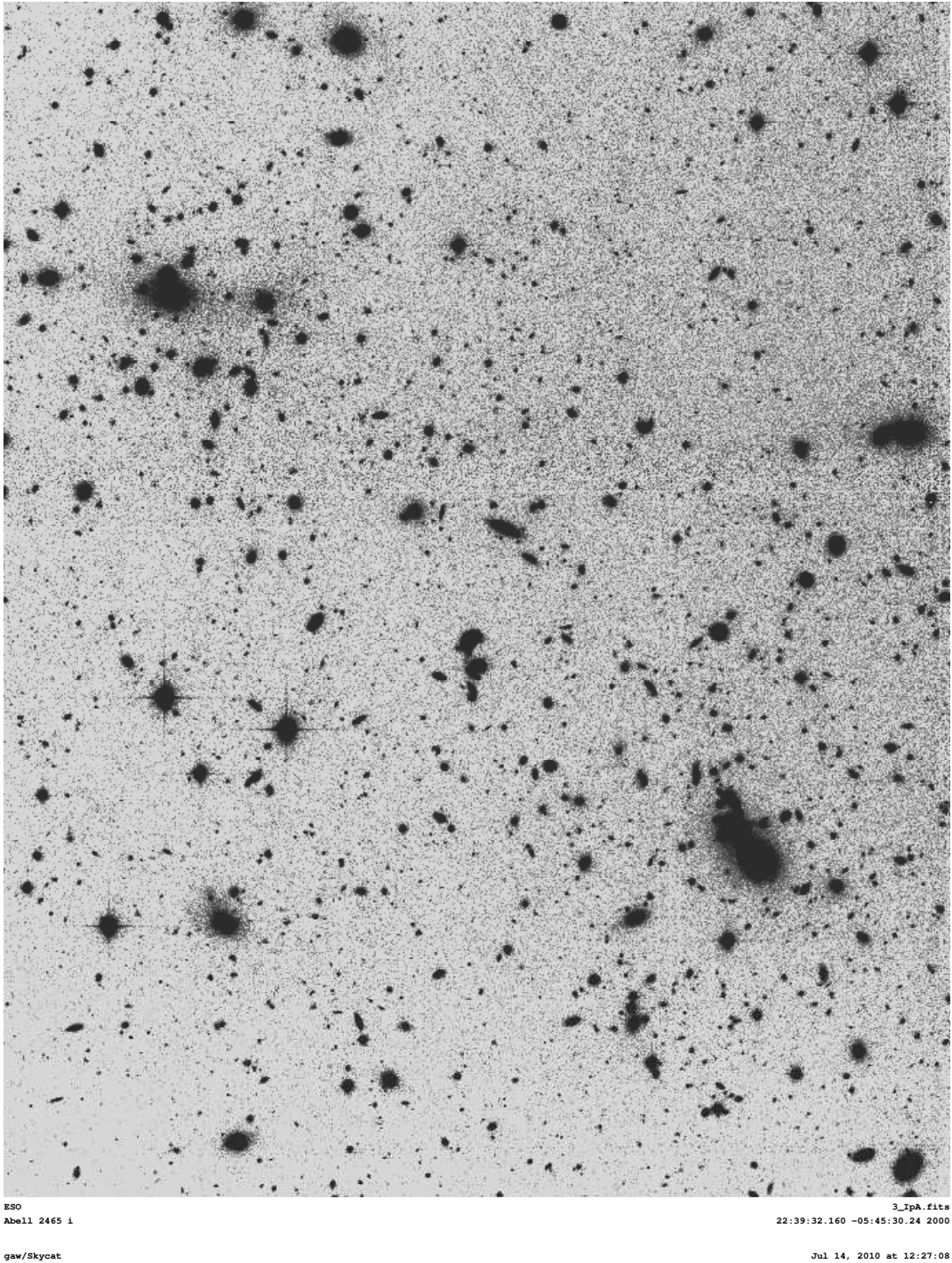}
\caption{A portion of the combined CFHT $i'$ Megaprime images
of the centre of Abell 2465 showing the NE (upper left) and SW (lower right)
clumps. The vertical edges of the picture are $8 \farcm 1$ in length.
North is to the top and East is to the left.
(Plot made using ESO's SkyCat, 
http://archive.eso.org/cms/tools-documentation/skycat).}
\label{cluster.eps}
\end{center}
\end{figure*}
%\newpage
%\clearpage

\section{OBSERVATIONS}
Figure~\ref{cluster.eps} shows the central $8 \farcm 1 \times 8 \farcm 1$ 
section of the $i'$ CFHT image
described below, containing both clumps of Abell 2465 (hereafter denoted
SW and NE). Basic astronomical data are given in Table 1.
The centres of the two clumps are separated by $5 \farcm 5$.

\subsection{Imaging Data}
The main part of the imaging data used in this paper is based on two sets of 
$r'$ and $i'$ images obtained by the QSO group of the CFHT in
2009. Five dithered $r'$ images of 300 seconds exposure each were taken
17 August at mean airmass 1.765 and five dithered $i'$ images, each  of 
412 seconds were observed 23 August at mean airmass 1.36 using the 
Megaprime instrument.\footnote{A description of this instrument 
including filters and {\sc Elixer} reductions can be
found on the CFHT web-page: 
http://ftp.cfht.hawaii.edu/Instruments/Imaging/MegaPrime/}
The {\sc Elixer} reductions of 
these images provided by the CFHT, which included bias subraction and
flattening, were  employed and the photometric zero
points and extinctions provided for the run were used, although these were
checked as explained below. The sets of dithered images were combined 
using the {\sc mscred} programs in {\sc IRAF}
\footnote{{\sc IRAF} is distributed by the National Optical Astronomy Observatories,
which are operated by the Association of Universities for Research
in Astronomy, Inc., under cooperative agreement with the National
Science Foundation.
}. The task {\sc mscfinder} 
was employed to
put coordinates to the WCS scale using the USNO-b catalogue. The program
{\sc mscimage} was used to make one single image from the 36 
individual CCD images
and each of these were stacked to make a final $r'$ and $i'$ image with
the task {\sc mscstack}. The resulting 
FWHM of stellar images in the vicinity of Abell
2465 are $0 \farcs 81$ and $0 \farcs 47$ for $r'$ and $i'$ respectively.

\subsubsection{Photometry}
The photometry of the images was measured using the {\sc Sextractor} programme 
(Bertin \& Arnouts 1996; Bertin 2009; Holwerda 2005). The single image 
mode was firstly used to scan the $i'$ image, which was secured 
in better seeing, to locate objects,  
and the double image mode was secondly run for the $r'$ image.
The {\sc MAG\_AUTO} option and mostly default parameters
recommended by Bertin (2009) for measurements of galaxies were set in the
Sextractor. The colour zeropoints and extinctions provided by the {\sc Elixer}
processing were employed. These were checked for the brightest galaxies
using CCD images of Abell 2465 obtained under photometric conditions
from the 1.3 m telescope at the MDM Observatory in Arizona with Kron-Cousins
$R$ and $I$ filters on the nights of 1995 November 18, 19, and 21 
and 2009 October 15 and 16 and calibrated
using Landolt (2009) with the result that $R = r' - 0.13 \pm 0.03$
and $I = i' -0.61 \pm 0.05$. Jordi \etal (2006) and Chionis \& Gaskell (2008)
give comparable values within the errors
for objects in the early-type galaxy colour range.
 
\subsubsection{Star/Galaxy Separation}
\begin{figure}
\includegraphics[width=8cm]{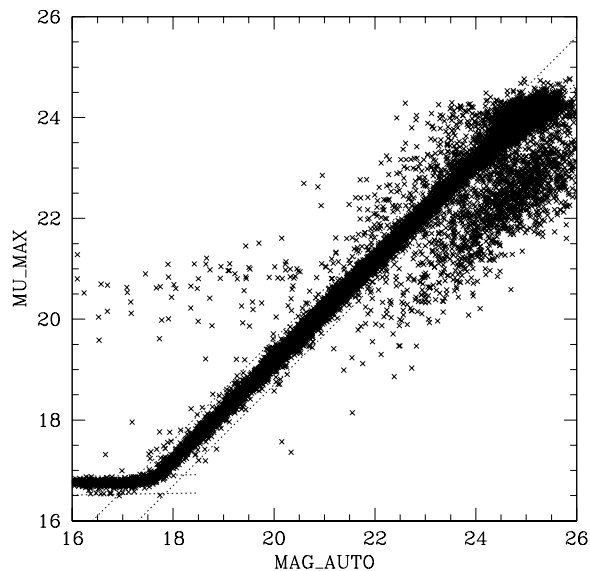}
\caption{{\sc Mu\_MAX} versus $i'$ {\sc MAG\_AUTO} from 
{\sc Sextractor} 
for stellar-like objects with $s > 0.8$ in the 
CFHT image. Dotted lines give the outlines of the
region used to reject stars.
}
\label{Mu_max.eps}
\end{figure}

The {\sc Sextractor} programme provides the {\sc CLASS\_ STAR}
stellarity parameter $0 \leq s \leq 1$, whereby
objects with $s \approx 1$ are stellar-like and $s \approx 0$ 
are galaxy-like. The $s$ grows increasingly imprecise for faint
sources due to seeing effects, so 
a better delineation between stars and galaxies is to employ the relation
between {\sc MAG\_AUTO}, the Kron-like elliptical aperture 
magnitude, and {\sc MU\_MAX}, the peak surface brightness above background
(Leauthaud \etal 2007; Penny \etal 2010). Figure~\ref{Mu_max.eps} 
plots objects in the $i'$ image for which $s \geq 0.8$. The stellar locus
and the outlined area used to determine which objects
were stars is shown. If an object lies in this region and 
has $s \geq 0.8$, it was classed as a star.

\begin{table}
 \caption{Basic Data for Abell 2465 used in This Paper}
\label{tab1objects}
\begin{center}
\leavevmode
\begin{tabular}{@{}ll}
\hline
(1) & (2) \\
\hline 
NE XMM $\alpha_{J2000}$ & 22 39 39.02\\
NE XMM $\delta_{J2000}$  &-05 43 28.2\\
  & \\
SW XMM $\alpha_{J2000}$&22 39 24.65\\
SW XMM $\delta_{J2000}$ &-05 47 15.0\\ 
  & \\
NE BCG $\alpha_{J2000}$&22 39 40.491\\
NE BCG $\delta_{J2000}$&-05 43 26.75\\
 & \\
SW BCG $\alpha_{J2000}$& 22 39 24.572\\
SW BCG $\delta_{J2000}$& -05 47 17.37\\
 & \\
Mean Redshift $z^1$ &0.2453 $\pm$ 0.0002\\
Luminosity Distance$^2$&  1224 Mpc \\
Angular-size Distance$^2$& 791 Mpc\\
Distance Modulus$^2$&40.44 mag.\\
Cosmology Corrected Scale$^2$& 230.06 kpc~arcmin$^{-1}$\\ 
Galactic Extinction $A_I^2$ &0.077 mag\\
K-term $K_I(z)^3$ & 0.15 mag\\
\hline
\end{tabular}
\end{center}
\medskip

{$^1$This paper, mean of 149 redshifts}
{$^2$From NED using the WMAP 5-year parameters. The NASA/IPAC Extragalactic
Database (NED) is operated by the Jet Propulsion Laboratory, California 
Institute of Technology, under contract with the National Aeronautics and 
Space Administration.}
$^3${Blanton \& Roweis (2007); Fukugita \etal (1995)}
\end{table} 

\subsubsection{The Red Sequence}
\begin{figure}
\includegraphics[width=0.5\textwidth]{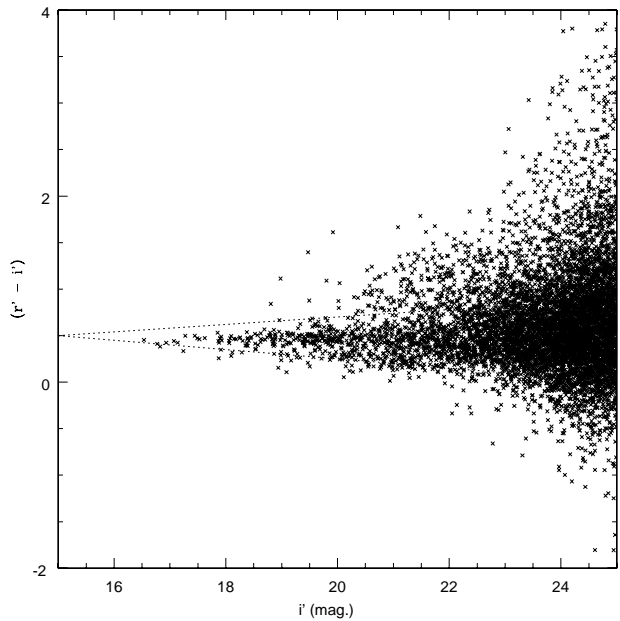}
\caption{The red sequence for the objects classed as
non-stellar in the inner $ 18 \arcmin \times 12 \arcmin$ central
portion of the CFHT images. Dotted lines give the outlines of the
region used to select red sequence cluster members.
}
\label{redseq.eps}
\end{figure}

Using the objects classified as non-stellar or hence galaxies discussed
above, the plot of $i'$ against $(r' - i')$ colours centred on Abell 2465
shows a well defined red sequence. Figure-\ref{redseq.eps} 
shows the magnitude-colour plot for the
inner $18 \arcmin \times 12 \arcmin$ rectangle which contains the two
subclusters. The red sequence is visible and the region used in the following 
is indicated. The chosen region extends to fainter magnitudes and is
widened to account for the increasing magnitude errors. 

\subsection{Spectroscopic Data}
A more detailed dynamical analysis of Abell 2465 will be presented eleswhere
as observations are still ongoing (Wegner \etal In preparation). Redshifts of
galaxies centred on Abell 2465 were measured
2007-2009 from the MDM Observatory and the Anglo-Australian Observatory
(AAT) before the CFHT images were obtained. 
The target list was constructed using $B$ and $R$ CCD images from
the MDM 1.3 telescope of the inner $10 \arcmin$ of the cluster
and the superCosmos Sky Survey (SSS), \footnote{http://www-wfau.roe.ac.uk/sss/index.html} which is based on the UK Schmidt $B_JRI$ photographic plates,
to cover the entire one degree square centred on the cluster. Sources flagged
as galaxies (i.e. {\sc CLASS} = 1), were
included. The zero points of the SSS photographic photometry were adjusted
using the CCD data and sources with colour indices, 
$1.5 \leq (B_J - R_2) \leq 2.4$ which
is centred on the red sequence were included in the list.

Redshifts of 359 galaxies were obtained.
Of these, 160 have redshifts in the range 70000 $< cz <$ 77000 \kms
which includes the cluster members. Of the remainder, 107 lie in the 
foreground and 92 lie behind the cluster. An additional 57 objects were
stars and 93 targets did not have high enough signal-to-noise to 
secure a redshift. 

The MDM observations were obtained with a long slit at the 
2.4 m Hiltner telescope.  The spectrograph was
rotated in order to acquire multiple galaxies, ranging in number typically 
from between three and ten objects, simultaneously. In 2007 and 2009, the
CCDS spectrograph with a 150 lines/mm grating was used. In 2007, the
slit width was $1 \farcs 3$ and the wavelength covered was 
$\lambda\lambda 4046-7245$. In 2009, a $2 \farcs 1$ slit and wavelength
coverage of $\lambda\lambda 4671-8591$ were used. These setups yielded
FWHM resolutions of 11 and 16 \AA, measured from night sky lines. In 2008,
the MKIII spectrograph, a $2 \farcs 36$ slit, and a 300 lines/mm grism 
covering $\lambda\lambda 4358-8716$ were employed, yielding a FWHM
resolution of 17 \AA.\footnote{Further details of these instruments can be
found on the MDM Observatory web-page:
http: //www.astro.lsa.umich.edu/obs/mdm/technical/index.html}
 Three 20 minute integrations were usually made with
interspersed wavelength calibrations. Standard data reductions of the CCD
spectra were carried out using {\sc IRAF}, which includes bias subtraction, flat 
fielding, and wavelength calibration. 

The largest number of redshifts was collected  through  the Service Observing
Programme of the AAT with the AAOmega multi-object spectrograph
in 2008 May 31 and July 25. Four 30 
minute exposures were obtained each night.
The AAOmega simultaneously observes the blue and 
red portions of the spectrum. For the blue, the 580V grating which covers
$\lambda\lambda 3700-5800$ was employed while for the red, the 385R grating
that extends across $\lambda\lambda 5600-8800$ was used. The instrument 
nominally has
392 fibres of $2 \farcs 0$ diameter and this setup gives a FWHM resolution of 
about 6 \AA. Wavelength calibration, flat fields, and biases were provided
and the preliminary reductions were facilitated using the {\sc 2dfdr} 
programme.\footnote{Descriptions of the AAOmega can be
found at: http://www.aao.gov.au/local/www/aaomega/}
Subsequent reductions were done with {\sc IRAF}.

\begin{table*}
\begin{minipage}{126mm}
\caption{Redshifts for cluster members in the central regions of Abell 2465 SW 
and NE}
%\hline
\begin{tabular}{lccrccccc}
\label{czdata}

$\alpha_{J2000}$ & $\delta_{J2000}$ & $cz$ (\kms)&$\varepsilon cz$ (\kms)&$N_{obs}$ & Tel.& $i'$ (mag.)& $w_i(SW)$& $w_i(NE)$\\
\hline
%\startdata
     339.75808 &       -5.79450 &   72990 & 29     & 1    & A    &  19.231 &0.6540 & 0.3450\\        
     339.78149 &       -5.75772 &   72540 & 196    & 1    & A    &  19.114 &0.6790 & 0.3200\\        
     339.79600 &       -5.79644 &   73972 & 37    & 1    &  M   &  18.894 &0.7900 & 0.2090\\        
     339.79696 &       -5.83281 &   72338 & 63    & 1    &  A   &  18.394 &0.7280 & 0.2710\\        
     339.80219 &       -5.78419 &   73571 & 177    & 1    & M    &  19.456 &0.8310 & 0.1690\\        
     339.80887 &       -5.74450 &   73949 & 9    & 1    & A    &  19.269 &0.7130 & 0.2860\\        
     339.81299 &       -5.75097 &   72992 & 35    & 1    & M    &  18.626 &0.7900 & 0.2090\\        
     339.81662 &       -5.79906 &   73586 & 192    & 1    & A    &  18.795 &0.9170 & 0.0820\\        
     339.82141 &       -5.70861 &   73235 & 139    & 1    & M    &  19.976 &0.5470 & 0.4520\\        
     339.82428 &       -5.74400 &   74204 & 108    & 1    & M    &  19.846 &0.7180 & 0.2810\\        
     339.82697 &       -5.74289 &   74196 & 45    & 1    & M    &  19.094 &0.7220 & 0.2780\\        
     339.83484 &       -5.82644 &   72171 & 213    & 1    & M    &  18.867 &0.8000 & 0.1990\\        
     339.83499 &       -5.73975 &   74032 & 115    & 1    & A    &  19.031 &0.7060 & 0.2940\\        
     339.83524 &       -5.75508 &   73520 & 110    & 1    & M    &  18.984 &0.8800 & 0.1190\\        
     339.83542 &       -5.82269 &   72566 & 10    & 1    & A    &  17.138 &0.9230 & 0.0760\\        
     339.84021 &       -5.82114 &   74470 & 58    & 2    & 2M    &  18.410 &0.8820 & 0.1170\\        
     339.84076 &       -5.80964 &   74604 & 113    & 1    & A    &  18.826 &0.8850 & 0.1140\\        
     339.84311 &       -5.75219 &   72357 & 150    & 1    & M    &  18.301 &0.8370 & 0.1620\\        
     339.84338 &       -5.79100 &   72788 & 69    & 2    &1A1M     &  18.886 &1.0000 & 0.0000\\        
     339.84662 &       -5.75625 &   73655 & 14    & 1    & M    &  18.652 &0.9140 & 0.0850\\        
     339.84725 &       -5.86158 &   72244 & 125    & 1    & M    &  18.225 &0.7260 & 0.2730\\        
     339.84738 &       -5.74164 &   74152 & 40    & 1    &  A   &  18.629 &0.6980 & 0.3010\\        
     339.84784 &       -5.81222 &   73476 & 30    & 1    & M    &  17.460 &1.0000 & 0.0000\\        
     339.85004 &       -5.74921 &   73575 & 93    & 2    & 2M    &  18.565 &0.7930 & 0.2060\\        
     339.85236 &       -5.78806 &   73533 & 1    & 5    & 2A3M    &  16.534 &1.0000 & 0.0000\\        
     339.85291 &       -5.78672 &   73522 & 126    & 1    & M    &  19.512 &1.0000 & 0.0000\\        
     339.85416 &       -5.78444 &   72242 & 14    &  2   & 2M    &  17.879 &1.0000 & 0.0000\\        
     339.85526 &       -5.78358 &   72094 & 171    & 1    & M    &  18.061 &1.0000 & 0.0000\\        
     339.85626 &       -5.78508 &   73561 & 113    & 1    & M    &  18.351 &1.0000 & 0.0000\\        
     339.85654 &       -5.78022 &   73429 & 30    & 1    &  M   &  18.370 &1.0000 & 0.0000\\        
     339.85669 &       -5.76196 &   72686 & 141    & 2    & 1A1M    &  18.528 &0.9430 & 0.0560\\        
     339.85796 &       -5.66972 &   72573 & 240    & 1    & M    &  18.873 &0.3360 & 0.6630\\        
     339.85922 &       -5.77811 &   72956 & 242    & 1    & A    &  19.912 &0.9610 & 0.0380\\        
     339.86432 &       -5.88822 &   73537 & 76    & 1    & M    &  18.776 &0.6480 & 0.3510\\        
     339.86508 &       -5.73889 &   73976 & 24    & 1    & A    &  20.007 &0.5240 & 0.4750\\        
     339.86609 &       -5.79450 &   73592 & 23    & 1    & M    &  17.487 &1.0000 & 0.0000\\        
     339.86728 &       -5.76581 &   72014 & 40    & 1    & A    &  18.565 &0.8410 & 0.1580\\        
     339.87332 &       -5.80614 &   73914 & 168    & 1    & A    &  19.535 &0.9330 & 0.0660\\        
     339.87396 &       -5.76278 &   74040 & 147    & 1    & M    &  19.918 &0.7570 & 0.2420\\        
     339.87462 &       -5.69311 &   73193 & 28    & 2    & 2A    &  19.127 &0.1460 & 0.8530\\        
     339.88004 &       -5.67217 &   73317 & 160    & 1    & M    &  19.517 &0.1810 & 0.8180\\        
     339.88092 &       -5.75026 &   74626 & 158    & 1    & A    &  17.853 &0.5010 & 0.4980\\        
     339.88297 &       -5.68617 &   72702 & 190    & 1    & M    &  19.378 &0.1810 & 0.8180\\        
     339.88324 &       -5.68225 &   73647 & 36    &  1   & M    &  18.848 &0.1070 & 0.8920\\        
     339.88464 &       -5.68631 &   73239 & 114    & 1    &M     &  19.236 &0.1010 & 0.8980\\        
     339.88464 &       -5.76603 &   73244 & 8    &   1  & M    &  18.565 &0.7930 & 0.2060\\        
     339.88483 &       -5.76292 &   73306 & 22    & 1    & A    &  19.680 &0.7460 & 0.2530\\        
     339.88586 &       -5.68911 &   73326 & 42    &  1   & M    &  21.910 &0.1390 & 0.8600\\        
     339.88846 &       -5.80075 &   74298 & 76    & 1    & M    &  19.315 &0.8170 & 0.1820\\        
     339.89209 &       -5.80678 &   73395 & 7    & 1    & M    &  20.391 &0.8510 & 0.1490\\        
     339.89420 &       -5.81267 &   73756 & 58    & 1    & M    &  18.386 &0.8430 & 0.1560\\        
     339.89771 &       -5.80597 &   74019 & 45    & 1    &A     &  19.407 &0.7810 & 0.2180\\        
     339.89868 &       -5.82036 &   73171 & 47    & 1    &M     &  19.533 &0.7860 & 0.2130\\        
     339.89879 &       -5.69512 &   75000 & 74    & 3    &1A2M     &  17.497 &0.0070 & 0.9920\\        
     339.89908 &       -5.82653 &   74029 & 44    & 1    & M    &  20.867 &0.7320 & 0.2670\\        
     339.89958 &       -5.70617 &   73388 & 179    & 2    & 1A1M    &  18.445 &0.0000 & 1.0000\\        
     339.90045 &       -5.83000 &   74071 & 78    & 1    & A    &  19.588 &0.7190 & 0.2800\\        
     339.90379 &       -5.75611 &   73388 & 130    & 1    & M    &  18.578 &0.3500 & 0.6490\\        
     339.90479 &       -5.74742 &   73297 & 56    & 1    & A    &  18.833 &0.1950 & 0.8050\\        
     339.90482 &       -5.67711 &   75208 & 54    & 1    & A    &  18.703 &0.1820 & 0.8170\\        
     339.90521 &       -5.68931 &   74709 & 51    & 1    & M    &  18.841 &0.0830 & 0.9160\\        
     339.90805 &       -5.72461 &   73311 & 126    & 1    & M    &  18.096 &0.0000 & 1.0000\\        
     339.90917 &       -5.77036 &   73698 & 13    & 1    & M    &  19.507 &0.5080 & 0.4910\\        
     339.90925 &       -5.77828 &   74085 & 35    & 1    & A    &  19.569 &0.5670 & 0.4320\\        

%\enddata
\end{tabular}
%\hline 
\end{minipage}
\end{table*}

\begin{table*}
\begin{minipage}{126mm}
\contcaption{Redshifts for cluster members in the central 
regions of Abell 2465 SW and NE}%\\
%\hline
\begin{tabular}{lccrccccc}
\label{czdata2}

$\alpha_{J2000}$ & $\delta_{J2000}$ & $cz$ (\kms)&$\varepsilon cz$ (\kms)&$N_{obs}$ & Tel.& $i'$ (mag.)& $w_i(SW)$& $w_i(NE)$\\

\hline
%\startdata
     339.90976 &       -5.75339 &   73857 & 80    & 1    & A    &  18.885 &0.2480 & 0.7510\\        
     339.90982 &       -5.73439 &   73220 & 185    & 1    & M    &  20.117 &0.0970 & 0.9020\\        
     339.91043 &       -5.69272 &   74636 & 48    & 1    & M    &  18.059 &0.0020 & 0.9970\\        
     339.91147 &       -5.81964 &   72128 & 64    & 1    & A    &  17.652 &0.7160 & 0.2830\\        
     339.91183 &       -5.68417 &   74958 & 4    & 2    & 2M    &  18.740 &0.1180 & 0.8810\\        
     339.91187 &       -5.63333 &   73217 & 115    & 1    & A    &  19.722 &0.2760 & 0.7230\\        
     339.91226 &       -5.83436 &   72025 & 40     & 1    & A    &  20.034 &0.6270 & 0.3720\\        
     339.91293 &       -5.79492 &   73180 & 37    & 1    & A    &  17.488 &0.6930 & 0.3060\\        
     339.91388 &       -5.73789 &   73060 & 23    & 1    & M    &  18.805 &0.0720 & 0.9270\\        
     339.91403 &       -5.73167 &   73361 & 76    & 1    & M    &  19.879 &0.0030 & 0.9960\\        
     339.91522 &       -5.73214 &   72576 & 40    & 3    &3M     &  18.076 &0.2090 & 0.7900\\        
     339.91864 &       -5.72389 &   73044 & 31    & 3    &1A2M     &  17.070 &0.0000 & 1.0000\\        
     339.91916 &       -5.72181 &   74061 & 119    & 2    &2M     &  17.898 &0.0000 & 1.0000\\        
     339.92084 &       -5.73189 &   74847 & 76    & 1    &M     &  19.318 &0.1950 & 0.8040\\        
     339.92203 &       -5.73417 &   74721 & 121    &1     &M     &  18.378 &0.1010 & 0.8980\\        
     339.92871 &       -5.74033 &   73186 & 93    & 1    & M    &  18.394 &0.0890 & 0.9100\\        
     339.92899 &       -5.70703 &   72670 & 142    & 1    & M    &  19.281 &0.0450 & 0.9540\\        
     339.93253 &       -5.72194 &   72745 & 74    & 2    &1A1M     &  17.960 &0.0000 & 1.0000\\        
     339.93542 &       -5.67039 &   73467 & 131    & 1    & M    &  19.377 &0.1130 & 0.8860\\        
     339.93896 &       -5.75525 &   73063 & 63    & 2    &1A1M     &  19.138 &0.2750 & 0.7240\\        
     339.94089 &       -5.71742 &   73814 & 21    & 1    & M    &  18.773 &0.0240 & 0.9750\\        
     339.94403 &       -5.65308 &   74188 & 150    & 1    & M    &  18.526 &0.2250 & 0.7740\\        
     339.94528 &       -5.76256 &   73828 & 42    & 1    & A    &  19.616 &0.3020 & 0.6970\\        
     339.94589 &       -5.64414 &   74096 & 264    & 1    & M    &  20.321 &0.3160 & 0.6830\\        
     339.94791 &       -5.62986 &   73698 & 77    & 1    & M    &  17.763 &0.2930 & 0.7060\\        
     339.95475 &       -5.79149 &   73571 & 138    & 3    &2A1M     &  19.103 &0.4570 & 0.5420\\        
     339.96658 &       -5.75661 &   73555 & 46    & 1    & A    &  17.582 &0.2830 & 0.7160\\        
     339.96689 &       -5.64408 &   73549 & 75    & 1    & A    &  20.273 &0.3160 & 0.6830\\        
     340.00000 &       -5.73803 &   73528 & 11    &  1   & A    &  18.061 &0.3430 & 0.6560\\        

%\enddata
\end{tabular}
%\hline 
\end{minipage}
\end{table*}

Absorption line measurements employed the Tonry \& Davis (1979)
cross-correlation method contained in
the {\sc IRAF} task {\sc fxcor}. KIII stars were initially used
for velocity standards with the MDM data, but for final reductions, the
BCG galaxies in Abell 2465 were used. These objects are in Table~\ref{czdata}
at $\alpha_{J2000}$ = 339.85236, 339.91864  and 339.91916. All spectra with
the Tonry \& Davis $R < 3$ and for which features could not be verified by
eye were rejected.

Emission lines were used when
an absorption line measurement could not be obtained. If both emission and
absorption velocities could be secured, they were averaged. The strong
H and K lines of Ca II were also used as a check. A substantial 
number of galaxies had emission lines in their spectra that could be
accurately measured and were used to verify the velocity scales of the
AAT data. Compared to the imaging data, all spectra are of bright
cluster members; the number of objects observed in the current set of spectra
drops off rapidly fainter than $i' = 20$ which is a rough estimate of the
limit of the current spectra. 

Table~\ref{czdata} presents a subset of the redshift data for cluster members
in the central regions of Abell 2465 which was used in this paper. 
Columns 1 and 2 are the galaxies' coordinates, 
Columns 3 and 4 are the measured heliocentric 
redshift and its error, Column 5 gives the
number of spectra, Column 6 states which telescope was used; A is for the
AAT and M is for MDM, Column 7 is the $i'$ magnitude from \S 2, and
Columns 8 and 9 are the fuzzy weights explained in \S 3.1 below.

\section{MASS ESTIMATES OF THE TWO CLUMPS}
An  estimate of the virial masses of the two clumps is made in the following. 
Although one can question the validity of this method,
Poole \etal (2006) find that colliding clusters regain virial equilibrium
relatively quickly.
Takizawa, Nagino, \& Matsushita (2010) employ N-body simulations to compare
virial mass estimates for colliding galaxy clusters and find
that when the mass ratio is larger than 0.25, the estimated virial masses can
be a factor two too large, and in general, X-ray mass estimates are more
accurate. While for the present, the question of virial equilibrium will be
avoided, one should note that the virial masses  
found in \S 3.3 agree within their errors with those obtained from 
X-ray scaling relations.

\subsection{Fuzzy Clustering}
A difficulty in dealing with galaxy clusters with overlapping components
is the separation of the members of each sub-clump, 
when not all the phase space information
can be known. For galaxy clusters with multiple components,
this becomes a problem when it is necessary to resolve the members of the
sub-clumps as in the present case of Abell 2465. This subject belongs to the
wider realm of cluster analysis (\eg Anderberg 1973; H\"{o}ppner
\etal 1999; Kaufman \& Rousseeuw 2005; Gan, Ma, \& Wu 2007) and many authors
have discussed solutions to the galaxy cluster problem, each with some
success in their particular case, but at present, no single method is
known to give a complete and secure solution. 
Pinkney \etal (1996) have reviewed several tests. 

Notable examples include the $\Delta$-statistic
(Dressler \& Schechtman 1988), the {\sc DEDICA} program of Ramella \etal 
(2007), and additional methods described in the papers of 
Salvador-Sol\'e, Sanrom\`a, \& 
Gonz\'alez-Casado (1993), Tully (1980), Serna (1996), and Diaferio 
(1999).  Tully (1980), Serna (1996), and Diaferio 
(1999)
employed the single-link hierarchical
clustering technique and constructed dendrograms. For the affinity parameter, 
Tully took the inverse of the attractive force, while 
Serna (1996) and Diaferio (1999) used the projected binding energy
between two galaxies, $i$ and $ j$:
\begin{equation}
E_{ij} = -G\frac{m_im_j}{|r_i-r_j|} +\frac{1}{2}\frac{m_im_j}{(m_i+m_j)}(v_i-v_j)^2,
\end{equation}
where $m_i, r_i,$ and $v_i$ are the mass, position on the sky, and the
redshift for the $i$th galaxy.

The k-medoid method (KMM) has been employed by several investigators 
(\eg Colless \& Dunn 1996; Kriessler \& Beers 1997; Yuan \etal 2005) 
to separate cluster members.
Kriessler \& Beers' KMM results compared favourably 
with previous
analyses of 56 clusters. The KMM assigns each object uniquely
to one cluster (Kaufman \& Rousseeuw 2005), termed
`hard clustering,' but given the observational errors and ambiguities
of the data, it is unlikely that such a unique assignment is always
accurate.

Consequently, the method of fuzzy analysis was explored
to separate the cluster
members of Abell 2465 (\eg Sato et al. 1997; H\"{o}ppner \etal 1999; Kaufman 
\& Rousseeuw 2005; Miyamoto \etal 2008). Fuzzy clustering 
generalizes the KMM and permits ambiguity in cluster membership by
providing a `membershift coefficient,' $w_i$, 
for each object, $i$, running from 0\%
for a non-member to 100\% for a member of only one cluster. Assigning an
object uniquely to a cluster by its largest membership coefficient 
(i.e. $w_i \ge 0.5$) returns the hard clustering result. 

The fuzzy analysis algorithm ({\sc FANNY}) and 
program of Kaufmann \& Rousseeuw (2005) were
employed for the analysis of the cluster members. As in the methods above, one 
chooses the number of clusters, $k$,  
an affinity between pairs of objects ($i,j$) and derives a 
dissimilarity matrix with elements $d(i,j)$. This was obtained by defining
a projected binding energy:
\begin{equation}
b_{ij} = \frac{-m_im_j}{|r_i-r_j|} + 1.162\times10^{-4} \mathcal{S} \frac{m_im_j}{(m_i+m_j)}(v_i-v_j)^2,
\end{equation}
using $r_i$ in megaparsecs, $v_i$ in \kms, 
and $m_i$ in units of 
$10^{12} M_\odot$. 
A scaling factor, $\mathcal{S},$ was employed to lower the weight of the 
second term. 
The $m_i$ were derived from the $i'$ magnitudes
of the galaxies using the formula of Cappellari \etal (2006) for the $I$-band,
$(M/L) = (2.35)(L_I/10^{10}L_{I,\odot})^{0.32}$ and $I = i' -0.61$. 
The $b_{ij}$ were converted to dissimilarities using
$d(i,j) = \frac{1}{2}[1+\rm{erf}(b_{ij}]$, 
which gives $0\le d(i,j) \le 1$ whereby nearby and tightly bound pairs with 
large negative $b_{ij}$ have small dissimilarities, and distant unrelated
pairs are assigned positive dissimilarities. 

\begin{figure}
\includegraphics[width=0.5\textwidth]{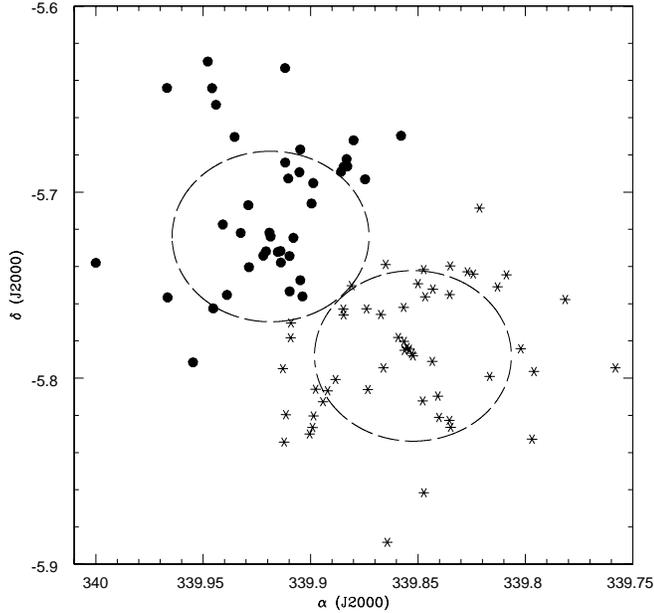}
\caption{Distributions of the galaxies in Abell 2465 with
measured redshifts and assigned to each clump
using the hard clustering (i.e. membership
coefficient, $w_i \ge 0.5$). Filled circles and
asterisks denote the NE and SW clumps respectively. The dashed circles are
centred on the BCG of each clump near the X-ray centres
and have radii of $2 \farcm 75$ or 0.63 Mpc which is half the distance
between the two centres.
}
\label{fan.eps}
\end{figure}

\begin{figure}
\includegraphics[width=1.0\textwidth]{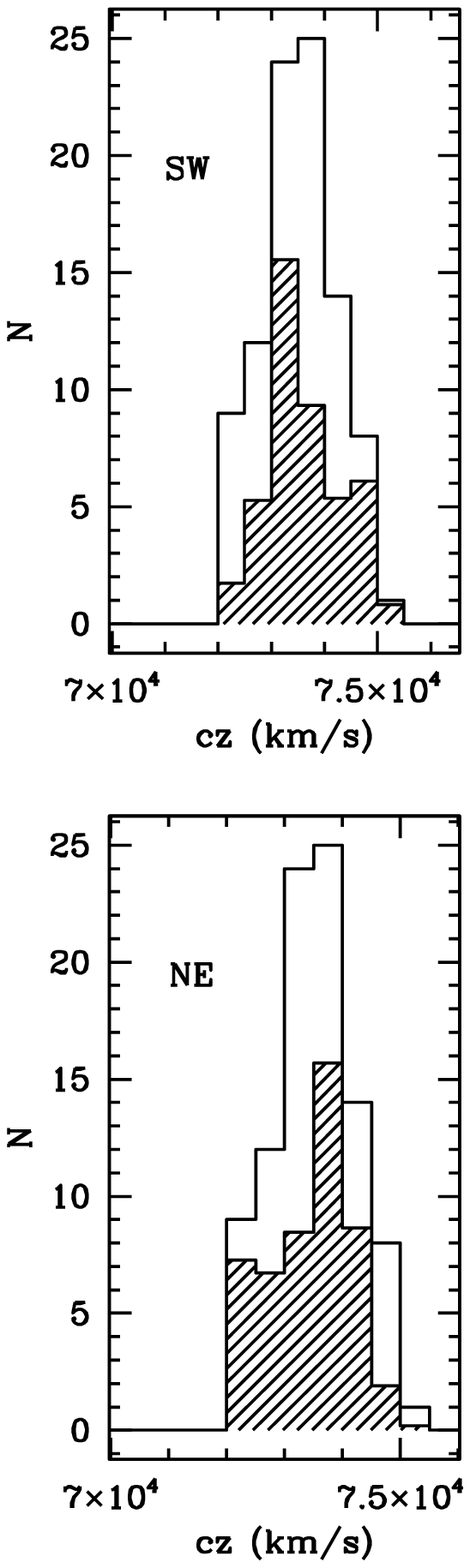}
\caption{Histograms of redshifts of the two Abell 2465
clumps based on the fuzzy clustering. Upper unfilled histogram shows the whole
sample. Lower shaded histograms show the weighted velocity distributions from
the fuzzy weights that were used in determininig the virial masses. 
}
\label{weightedvhist.eps}
\end{figure}

Figure~\ref{fan.eps} shows the resulting distribution of the galaxies in the 
two clumps and Figure~\ref{weightedvhist.eps} has histograms of their 
corresponding velocities. 
In this hard clustering presentation, the result is 
approximately what one intuitively expects as the galaxies are divided 
mostly into the two groups near the outlines of the circles in 
Figure~\ref{fan.eps} and using the fuzzy weights in 
Figure~\ref{weightedvhist.eps} the difference in the
velocity peaks can be seen.

\begin{figure*}
\centering
 \includegraphics[width=0.5\textwidth]{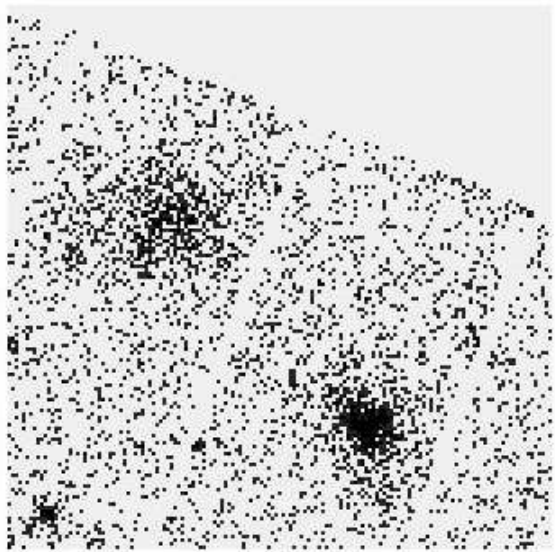}
 \includegraphics[width=0.45\textwidth]{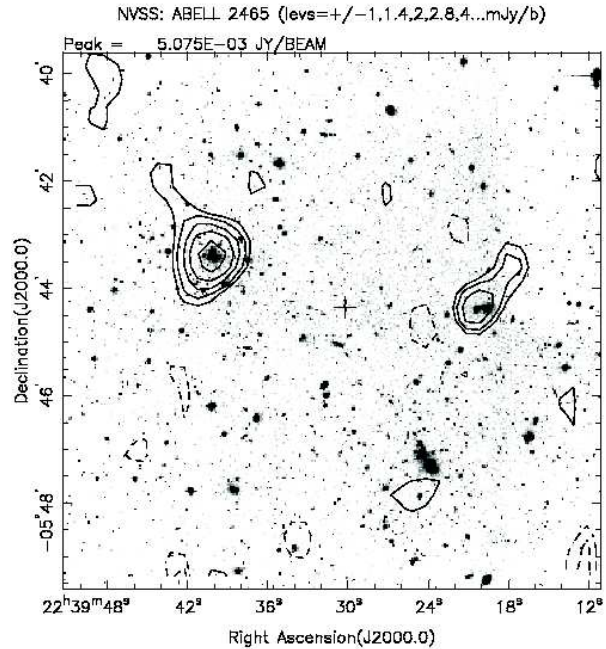}
\caption{The X-ray and 1.4 GHz radio images of Abell 2465. Both
are at the same scale. The vertical border of each box represents 2.3 Mpc at
the distance of Abell 2465. North is at the top and East is to the left.
(Left) 1-2 keV  X-ray data near the cluster from a portion of archived XMM 
image 0149410401003 from project of
PI S. Mathur. (Right) 1.4 GHz contours provided by the NVSS Postage 
Stamp Server superimposed on the $i'$ image of this paper.
}
\label{XMMNVSS}
\end{figure*}

\subsection{The Virial Masses}
Virial masses were estimated from the redshift data following  
\eg Carlberg \etal (1996) and Girardi \etal (1998).
All quantities were referred to the rest frame of the clusters.
A weighted virial mass estimator was used which reduces to the 
Heisler, Tremaine, \& Bahcall (1985) eq. 4 for unit weights:
\begin{equation}
M_{v} = \frac{3\pi \sum_iw_i}{2G}\frac{\sum_i w_iV_{zi}^2}{\sum_{i<j}w_iw_j/R_{\bot,ij}},
\end{equation}
where the $w_i$ are the fuzzy clustering weights, 
$V_{zi} = (V_i - \bar{V})/(1+\bar{z})$ and $R_{\bot,ij}$ is the angular
size distance between two galaxies $i$ and $j$. 

Galaxies within 1.4 Mpc of the clump centres and with redshift
velocities $72000 < V_{zi} < 76000$ \kms were used to obtain
$M_{v}$. 
The galaxies with their redshifts that were used are given in 
Table~\ref{czdata}.

For the
NE clump, there are $\sum_i w_i = 49$ galaxies within this radius 
while for the SW clump,
$\sum_i w_i = 44$. The resulting virial masses, $M_v$, for NE and SW 
respectively are:
%\[ 4.1 \pm 0.8 \times 10^{14} M_\odot} ~{\rm and}~ 3.8 \pm 0.8 \times 10^{14} M_\odot \]
where the uncertainties
are jackknife errors. Figure~\ref{weightedvhist.eps} shows the histograms of 
the velocities. The corresponding mean redshifts for the two clumps are 
$\bar{V}_{NE} = 73593 \pm 102$ \kms and
$\bar{V}_{SW} = 73388 \pm 109$ \kms, which yields a velocity
difference of $\Delta V = 205 \pm 149$  \kms.

The virial radii, according to the formula
$r_{200}=\frac{\sqrt{10}}{3}\frac{\sigma^2}{H(z)^2}$ (Carlberg \etal 1996)
are $1.21 \pm 0.11$ Mpc (NE) and $1.24 \pm 0.09$ Mpc (SW).
This assumes the virial mass is approximated by
$M_{200} = \frac{4}{3}\pi r_{200}^3 \Delta_c\rho_o(z)$,
where $\rho_o(z)$ is the critical density at redshift $z$ and
$\Delta_c$ is the cluster's density enhancement, set equal to 200.

The masses of the two clumps calculated using 
the hard clustering weighting ($w_i = 0~\rm{or}~1$) are 3.7 $\pm$ 0.7 and
3.0 $\pm$ 0.7 $\times 10^{14} M_\odot$ respectively for the SW and the NE
clumps. 
For comparison, the redshift difference between the two BCGs
near the centres of the NE and SW clump is 489 \kms. The unwighted average
of 149 cluster members inside one degree of the cluster centre is
73530 $\pm$ 58 \kms.

A correction to $M_{v}$
is required due to the whole cluster not being included in the calculation
(\eg Girardi \etal (1998 and references therein).
This depends on the galaxy and velocity dispersion distributions with radius. 
A $c = 6$ NFW profile was assumed and it was found that the correction could
be neglected for the present data.

\subsection{Comparison with X-ray Masses}

The virial masses can be compared with those from X-ray scaling relations. 
Both ROSAT and XMM-Newton observed Abell 2465 serendipitously. Contours of
the ROSAT data are in Perlman \etal (2002). Figure~\ref{XMMNVSS} shows the XMM
image alongside the 1.4 GHz radio data discussed in \S 3.4.
Averaging the 
unabsorbed ROSAT values in the (0.5-2.0) keV band according to Vikhlinin \etal
(1998) and Perlman \etal (2002) gives 
$f_X = 3.605 \times 10^{-13}$ 
and $2.53 \times 10^{-13}$ ergs~sec$^{-1}$cm$^{-2}$
respectively for the NE and SW clumps. The XMM-Newton values taken from the 
2XMMi\_DR3 catalogue in the (0.5-2.0) 
keV band are the ep\_2 +  ep\_3
fluxes and are $2.44 \times 10^{-13}$ and $2.08 \times 10^{-13}$
respectively. These were multiplied by 1.07 to correct for absorption, using 
$n_H = 3.64 \times 10^{20}$ atoms~cm$^{-2}$, (the mean of the 
Kalberla \etal (2005) and Dickey \& Lockman (1990) values implemented in
HEASARC), Wilms \etal (2000) for the X-ray absorptivity per H atom in the ISM, 
and a 4 keV Raymond-Smith (Raymond \& 
Smith 1977) 0.3 solar metals model at redshift $z = 0.245$.

Many authorities find scaling relations between X-ray luminosity
and mass, temperature, etc.
(\eg Reiprich \& B\"{o}hringer 2002; Popesso \etal 2005; Rykoff \etal
2008). The results of Popesso \etal
(2005) for $M_{200}$, in the (0.1-2.4 keV)  band were adopted.
A multiplicative factor of 1.60 was found to bring the ROSAT and XMM
(0.5-2.0) keV data into this band using the 4 keV Raymond-Smith model above.
With the luminosity distance of 1224 Mpc for 
Abell 2465, $L_X = 8.94 \pm 1.44 \times 10^{43}$
ergs~sec$^{-1}$ for NE and 
$L_X = 6.84 \pm 0.43 \times 10^{43}$ ergs~sec$^{-1}$ for SW.
The Popesso et al. (2005) uncorrected relation for mass, 
$\log M_{200} = [\log(L_X/10^{44})+1.15]/1.58$ yields:
\[ 4.4 \pm 0.6 \times 10^{14} M_{\odot} ~\rm{and}~ 3.6 \pm 0.2 \times 10^{14} M_\odot \]
for the NE and SW clumps respectively.

The $L_X - T$ relation,
$\log T_X = [\log(L_X/10^{44})+2.06]/3.30$ (Popesso \etal 2005), gives 
temperatures for NE and SW of:
\[ 4.1 \pm 0.3 ~\rm{keV} ~\rm{and}~ 3.75 \pm 0.2 ~\rm{keV}, \]
where the errors include the range in $L_X$ and the scatter of the
relation. The mass weighted $M_{200}-T_X$ relation found by Sanderson \etal
(2003) gives nearly identical values of $T_X$.

Given $T_X$, the
cooling times of the two clumps provide additional information.
The bremsstrahlung cooling time for cluster
gas of temperature T and hydrogen density, $n_p$ is:
\begin{equation}
t_{cool}=8.5\times10^{10}\bigg(\frac{T}{10^8}\bigg)^{\frac{1}{2}}\bigg(\frac{n_p}{10^{-3}\rm{cm}^{-3}}\bigg)^{-1} 
\end{equation}
(Sarazin 1986; eq. 5.23). At the current temperature of $T \approx 4$ keV, 
$n_p$ can be estimated from
\begin{equation}
\varepsilon = 3.0\times 10^{-27}\bigg(T_g\bigg)^{\frac{1}{2}}n_p^2,
\end{equation}
Sarazin (1986; eq. 5.21). 
Using that $L/2 \approx \varepsilon \frac{4}{3}\pi r_c^3$
and the core radii $r_c = 42$ and 130 kpc for SW and
NE (Vikhlinin \etal 1998),
$t_{cool} \approx 4$ and 20 Gyr for SW and NE respectively. This 
indicates that SW is a cooling core (CC) cluster.
This shorter $t_{cool}$ in the SW subcluster results from its smaller
$r_c$ and is consistent with studies of CC clusters showing
they have core radii $r_c \lse 100$ kpc (\eg
Chen \etal 2007). As already noted, there is no evidence for strong AGN
activity in either member of Abell 2465. It is tempting to identify the CC as 
a result of the cluster's merger, but O'Hara \etal (2006) argued that 
major mergers do not evolve cooling cores from their study of the scatter in 
scaling relations. However, ZuHone \& Markevitch (2009) found that 'sloshing'
produced in mergers could be a source of heating in cluster cores. 

\subsection{Radio data}
Abell 2465 appears to be detected in the 1.4 GHz NRAO VLA Sky Survey (NVSS; 
Condon \etal 1998). 
The radio contours are shown in Figure~\ref{XMMNVSS} where they are 
superimposed on the $i'$ image. A
source with peak flux $6.2 \pm 0.6$ mJy falls near the NE component and
appears to be a radio halo. A
second elongated object with a peak flux of $3.1 \pm 0.4$ mJy is 
near the three early-type cluster members located about 3 arcmin north of
the SW clump. 
No significant source lies in the SW optical component. If the two radio 
sources are at the distance of Abell 2465, 
their luminosities are $11 \times 10^{23}$ and $6 \times 10^{23}$ W Hz$^{-1}$
These are within the range of luminosity, temperature, and size for
diffuse radio halos and relics summarized by Feretti (2002). The radio
halo is centred near the NE subcluster which might identify it as the
primary component of Abell 2465. 

A more detailed analysis of the X-ray and radio data will be given elsewhere
(Wegner \& Johnson, in preparation). 

\subsection{Cluster Velocity Dispersion Measurements}
\begin{figure}
\centering
\includegraphics[width=0.9\textwidth]{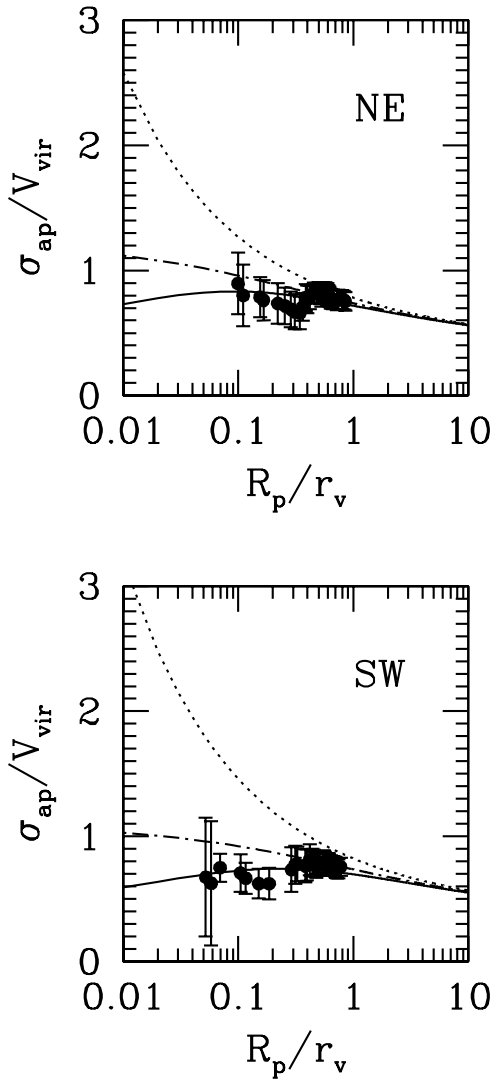}
\caption{Radial profiles of the aperture-velocity dispersions
for the two clumps in Abell 2465. The abscissae give the projected 
radial distances relative to the virial radius, $r_v$, and the ordinates
are the velocity dispersions and their errors
relative to the circular velocity, $V_{vir}$,
defined in the text inside of $R_p$. The different curves are for
NFW models with $c = 10$ (SW) and $c = 4$ computed 
using {\L}okas \& Mamom (2001). Dotted curve: $\beta = 1$, dot-dash
curve denotes $\beta = 0.5$, and solid curve is $\beta = 0$.
}
\label{sigprof.eps}
\end{figure}

The run of radial velocity dispersion with radius contains information on the 
dynamics of the clusters. The results of measurements for the two clumps
are shown in Figure~\ref{sigprof.eps}, where the fuzzy weighting was employed
and errors given are jackknife errors. 

The simplest case solves the spherical Jeans
equation for the anisotropy parameter, $\beta$, where $\beta = 0$ for
isotropic orbits, and $\beta = 1$ is
the non-physical limiting case of purely radial orbits. 
The aperture values of $\sigma$ are 
shown in Figure~\ref{sigprof.eps}, where the formulae of  {\L}okas
\& Mamon (2001) were used for the mean velocity dispersion 
inside an aperture of radius $R$:
\begin{equation}
\sigma_{ap}^2(R) = \frac{S^2(R)}{M_p(R)},
\end{equation}
where $M_p$ is described in equation (8) and $S^2$ is:
\begin{eqnarray}
S^2(R) = c^2g(c)M_v\bigg\{\int^\infty_0 \frac{\sigma^2_r(s,\beta)s}{(1+cs)^2}\bigg(1-\frac{2\beta}{3}\bigg)ds\nonumber\\
+\int^\infty_R\frac{\sigma^2_r(s,\beta)(s^2-\tilde{R}^2)^{1/2}}{(1+cs)^2}\bigg[\frac{\beta(\tilde{R}^2+2s^2)}{3s^2}-1\bigg]ds\bigg\},
\end{eqnarray}

\noindent where $s = r/r_v$, $M_v = \frac{4}{3}\pi r_v^3 \Delta_c \rho_o(z)$ 
is the virial mass, 
and the other quantities are defined in \S 5. {\L}okas \& Mamon 
(2001) give analytical expressions for ${\sigma^2_r(s,\beta)/V^2_v}$
for constant $\beta =$ 0, 0.5, and 1 and the Osipkov-Merritt (OM) model
(Osipkov 1979; Merritt 1985), $\beta_{OM} = s^2/(s^2+s_a^2)$, 
which are used to evaluate the integral. 

Figure~\ref{sigprof.eps} compares curves for three 
values of $\beta$
with the observed values in terms of the virial radius circular
velocity $V_{vir}^2=GM_v/r_v$. OM models with $s_a \approx \frac{4}{3}$
lie nearby the lower $\beta$ curves and are not shown. Values of $V_{vir}$
of 763 \kms and 722 \kms were adopted for the NE and SW clump respectively.
It can be seen that the most consistent anisotropy
values are low, near $\beta = 0$, but agreement with $\beta = 1.0$ is ruled
out for small radii. 
These values of $V_{vir}$ place the data points onto the 0 and 0.5 $\beta$
curves at the largest observed radii. Using the masses and
radii found above, the corresponding values would be 1077 and 1043 \kms,
but given the error in the value of $M_v/r_v$ which give $V_{vir}$, these
differ from the adopted values less than two standard deviations.

The behavior of $\beta$ in the dark matter component of galaxy clusters has 
been investigated by several authors with the idea that this can probe its
properties. For $\Lambda$CDM, Thomas \etal (1998) found that $\beta \leq 0.3$
inside the virial radius. Hansen \& Piffaretti (2007) obtained similar
values for two clusters as did Host (2009) and Host \etal (2009) for several
X-ray clusters. The values of $\beta$ for the subclusters of Abell 2465 are
consistent with these results.

\subsection{Emission Line Galaxies in Abell 2465}
Of the 149 galaxies in Abell 2465, observed spectroscopically with 
redshift $70000 \le cz \le 76000$ \kms, and within
one degree of the cluster centres, 55, or 37\%, show
detectable H$\alpha$ emission of which 38 also have
measureable [N II], H$\beta$, and [O III] and can be plotted in the
diagnostic diagram that separates star forming galaxies, liners, and AGNs. 
Equivalent widths were measured using the 
{\sc IRAF} {\sc splot} routine and
independently measured twice to estimate errors. The
H$\alpha$ and H$\beta$ equivalent widths were adjusted for
underlying absorption. Following Wegner \& Grogin (2008).
H$\alpha$ was corrected by adding 2.32 \AA ~to the measured emission and
2.02 \AA ~was added to the H$\beta$ emission, which are the 
absorption equivalent widths
of these lines from galaxies that appear free of emission.

The resulting line ratios and their errors are given in 
Table~\ref{emdata}, where
Columns 1 and 2 are the coordinates of the galaxies. Columns 2, 3, 4, and
5 give the line ratio measurements and their errors. In Column 6
A and M denote whether AAT or MDM spectra were measured, and the last column
is an estimated morphological type.

The morphological types of the emission line galaxies were estimated from the
$i'$ image. The largest proportion of these appeared to be disturbed
single objects without a visible companion. At least 24 or 44\% have
unusually asymmetrical disks or spiral arms. Only about six of these galaxies
have obvious companions or tidal tails.

The positions of the galaxies that have detected H$\alpha$ emission, 
but the other emission
lines were too faint to place them in the diagnostic diagram, are listed in
Table~\ref{emdata2} along with their morphologies.

\begin{figure} 
\includegraphics[width=0.45\textwidth]{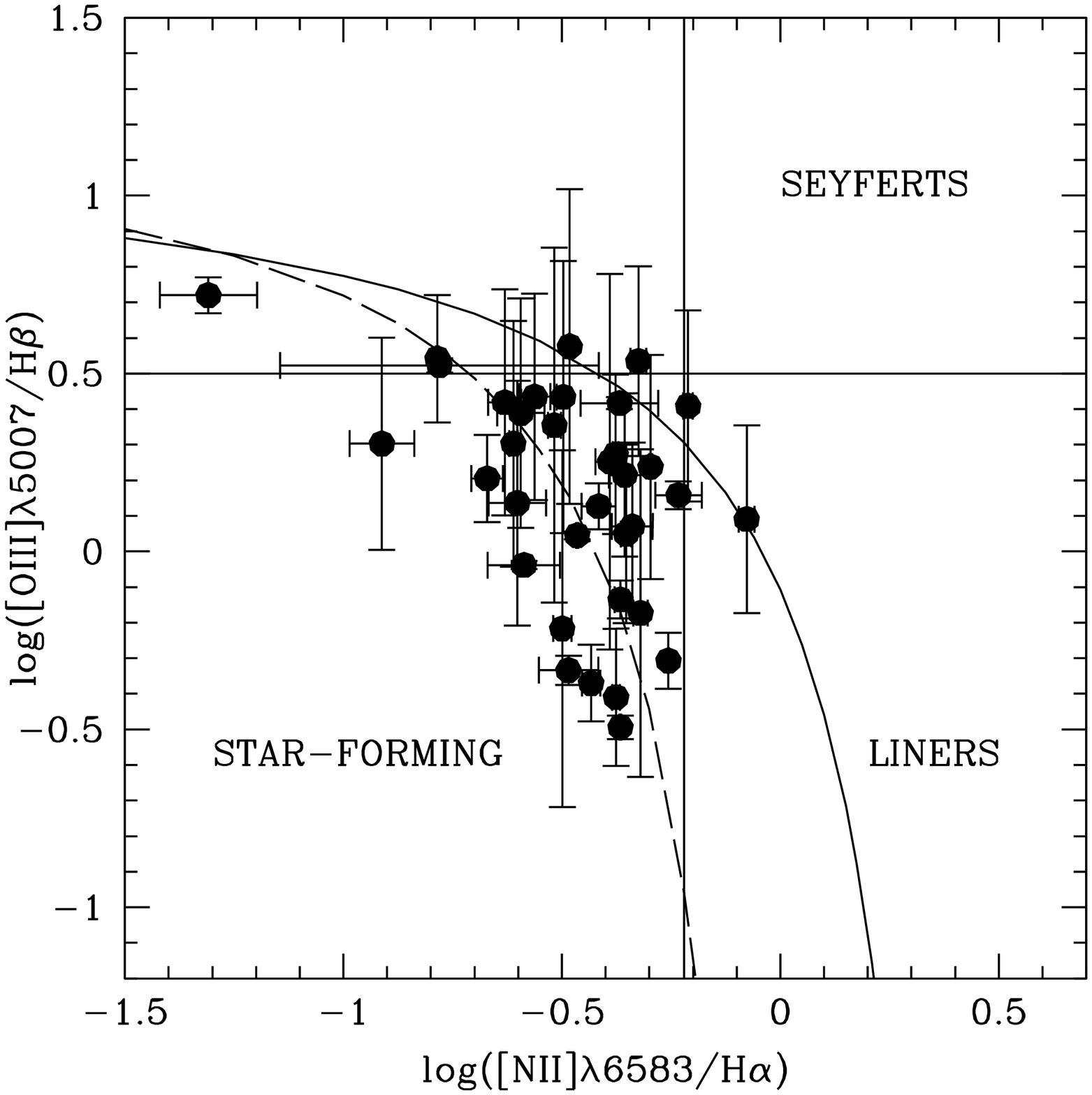}
\caption{The diagnostic diagram for the emission line galaxies
in Figure~\ref{emgalpos}. 
}
\label{diagnostic.eps}
\end{figure}

Figure~\ref{diagnostic.eps} shows the diagnostic diagram. The 
line separating the 
Seyfert and liner regions is that given in Yan et al. (2006). The
solid and dashed curves demarcating the
edge of the star-forming region are from Kewley et al. (2001) and
Kaufmann et al. (2003). The emission line galaxies in
Abell 2465 are nearly all star-forming objects. Only four objects lie on the
border of the liner or Seyfert region. 
There are no no definite Seyferts or AGN activity dominating either clump.

\begin{table*}
\begin{minipage}{126mm}
\caption{Line ratio data for the emission line galaxies shown in 
Figure~\ref{diagnostic.eps}}%\\
\label{emdata}
\begin{tabular}{lccccccl}
\hline

$\alpha_{J2000}$ & 
$\delta_{J2000}$ & 
$\log$[NII]/H$\alpha$ &
$\varepsilon \log$[NII]/H$\alpha$ &
$\log$[OIII]/H$\beta$ & 
$\varepsilon\log$[OIII]/H$\beta$&
Tel.& Type \\
\hline
%\startdata
     339.45346 &      -6.18531  & -0.499    & 0.021     & -0.217    & 0.501  & A & Sp\\   
     339.47983 &      -6.20242  & -0.562    & 0.036     & 0.435    & 0.290   & A & S0 pec \\   
     339.64358 &      -5.61342  & -0.320    & 0.016     & -0.173    & 0.460  & A & Sa asy \\   
     339.64612 &      -5.74183  & -0.353    & 0.012     & 0.049     & 0.251  & A & S0 pec \\   
     339.65378 &      -6.16689  & -0.077     & 0.018     & 0.091     & 0.264 & A & S0 \\   
     339.68289 &      -5.83781  & -0.376   & 0.009      & -0.410    & 0.193  & A & S0 pec \\   
     339.68362 &      -5.92178  & -0.631    & 0.038     & 0.419   & 0.318    & A & S0 \\   
     339.68655 &      -5.88831  & -0.416   & 0.039     & 0.127    &    0.064 & A & S0 asy+S0 \\   
     339.69240 &      -5.92532  & -0.671    &0.036      & 0.205     & 0.122  & 2A & Sc \\     
     339.74338 &      -5.94525  & -0.587    & 0.083     & -0.038     & 0.011 & A & Sp asy \\   
     339.75180 &      -5.98858  & -0.297    & 0.003      & 0.238     & 0.315 & A & SB0 rings \\   
     339.75809 &      -5.79450  & -0.433    & 0.021     & -0.370    & 0.108  & A & Sa asy \\   
     339.76382 &      -5.97494  & -0.356   & 0.036     & 0.215      & 0.229  & A & SBc asy \\   
     339.78741 &      -5.62375  & -0.785    & 0.003      & 0.542    & 0.179  & A & Sp edge on \\   
     339.80887 &      -5.74450  & -0.482    & 0.001       & 0.576    & 0.442 & A & Sa \\ 
     339.81514 &      -5.89681  & -0.368    & 0.089    & 0.417    & 0.017    & MA & S0 \\       
     339.82657 &      -5.40069  & -0.390    & 0.033    & 0.252   & 0.528     & A & Sa asy \\   
     339.83768 &      -5.99783  & -0.594    & 0.054     & 0.389      & 0.322 & A & S0 (POSS2) \\   
     339.84326 &      -5.75222  & -0.339   & 0.047     & 0.070     & 0.236  & A & SB0 \\   
     339.85074 &      -6.20175  & -1.309   & 0.111    & 0.720    &     0.051  & A & Sc \\   
     339.85663 &      -5.76189  & -0.256    & 0.001       & -0.307   & 0.079 & A & S0 asy \\   
     339.85699 &      -5.89256  & -0.602    & 0.066     & 0.136    & 0.344  & M  & Sa \\   
     339.86508 &      -5.73889  & -0.377    & 0.007      & 0.273    & 0.224  & A & S0+companion \\   
     339.87332 &      -5.80614  & -0.211    & 0.011     & 0.409    &  0.269  & A & Sp asy \\   
     339.88025 &      -5.56244  & -0.611    & 0.010     & 0.303    & 0.345   & A & E+Sp \\   
     339.88465 &      -5.68619  & -0.780   & 0.365    & 0.523    & 0.020    & 2M & Sp+E \\
     339.88483 &      -5.76292  & -0.297    & 0.013     & 0.018     & 0.326  & A & SBc \\   
     339.88586 &      -5.68911  & -0.912    & 0.074     & 0.303    & 0.298   & M & E? \\   
     339.89172 &      -5.60164  & -0.517   & 0.015    & 0.355    & 0.499     & A & Sb pec \\   
     339.89771 &      -5.80597  & -0.366    & 0.013     & -0.135    & 0.053  & A & Sp asy \\   
     339.90045 &      -5.83003  & -0.325    & 0.018     & 0.535    & 0.267   & M & S0 \\
     339.90931 &      -5.77829  & -0.703   & 0.073      & 0.283    & 0.165 & AM & Sp tail \\
     339.94437 &      -5.65339  & -0.534    & 0.016     & 0.200   & 0.377    & M & S0 pec \\   
     339.94528 &      -5.76256  & -0.233    & 0.053     & 0.158    &  0.039  & A & Sp asy \\   
     340.06308 &      -5.58347  & -0.485    & 0.068     & -0.334    & 0.041  & A & Sb asy \\   
     340.06516 &      -5.62183  & -0.465    & 0.002      & 0.045     & 0.011 & A & Sa asy \\   
     340.10855 &      -5.57439  & -0.497    & 0.015     & 0.434    & 0.382   & A & Sa asy \\   
     340.12149 &      -5.95319  & -0.366    & 0.002      & -0.494   & 0.033  & A & Sb asy \\   

%\enddata
\hline
\end{tabular}
\medskip

{NOTES TO THE TABLE: asy - one side of object noticeably stronger; 
pec - disturbed and/or shells }\\
\end{minipage}
\end{table*}

\begin{table}
\caption{Additional galaxies with detected H$\alpha$ emission
}%\\
\label{emdata2}
\begin{center}
\leavevmode
\begin{tabular}{lccl}
\hline

$\alpha_{J2000}$ & 
$\delta_{J2000}$ & 

Tel.& Type \\
\hline
%\startdata
 339.58828 & -5.93192 & A & E\\
 339.76129 & -5.87131 & A & S0 pec\\
 339.79599 & -5.79644 & M & Sp tail\\
 339.82037 & -5.85808 & M & S0 asy\\
 339.84738 & -5.74164 & A & E \\
 339.85510 & -5.78122 & M & SBc asy\\
 339.86731 & -5.76581 & A & E \\
 339.88293 & -5.68617 & M & S0\\
 339.88461 & -5.76603 & M & S0\\
 339.88842 & -5.80075 & M & S0\\
 339.90484 & -5.74722 & A & S0 \\
 339.92083 & -5.73192 & M & E \\
 339.93524 & -5.67008 & A & S0 \\
 339.93899 & -5.67408 & M & Sa \\
 339.94891 & -5.62542 & M & E \\
 340.10977 & -5.80497 & A & Sp\\

%\enddata
\hline
\end{tabular}
\end{center}
\end{table}

\begin{figure*} 
\centering
 \includegraphics[width=0.45\textwidth]{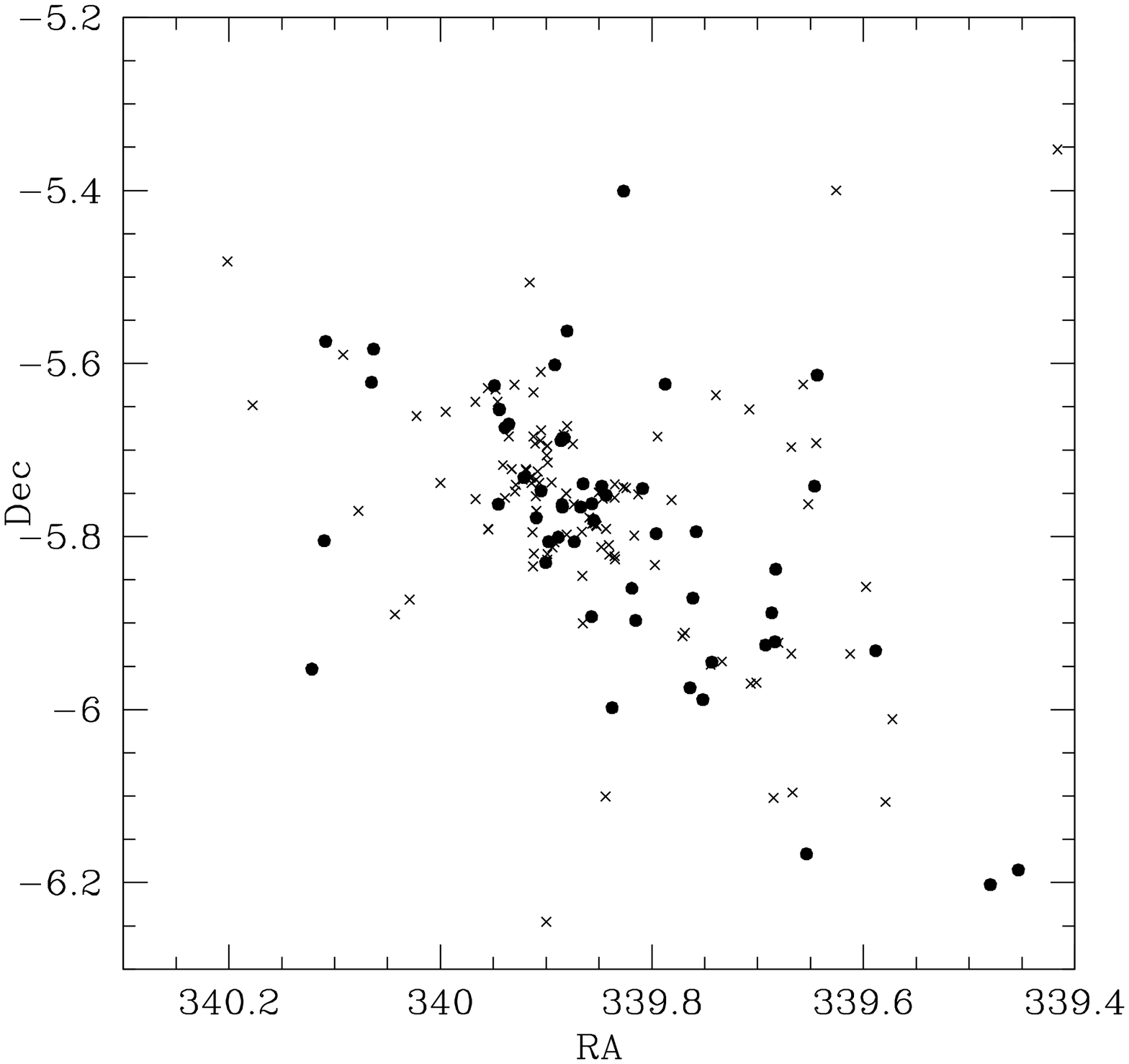}
 \includegraphics[width=0.45\textwidth]{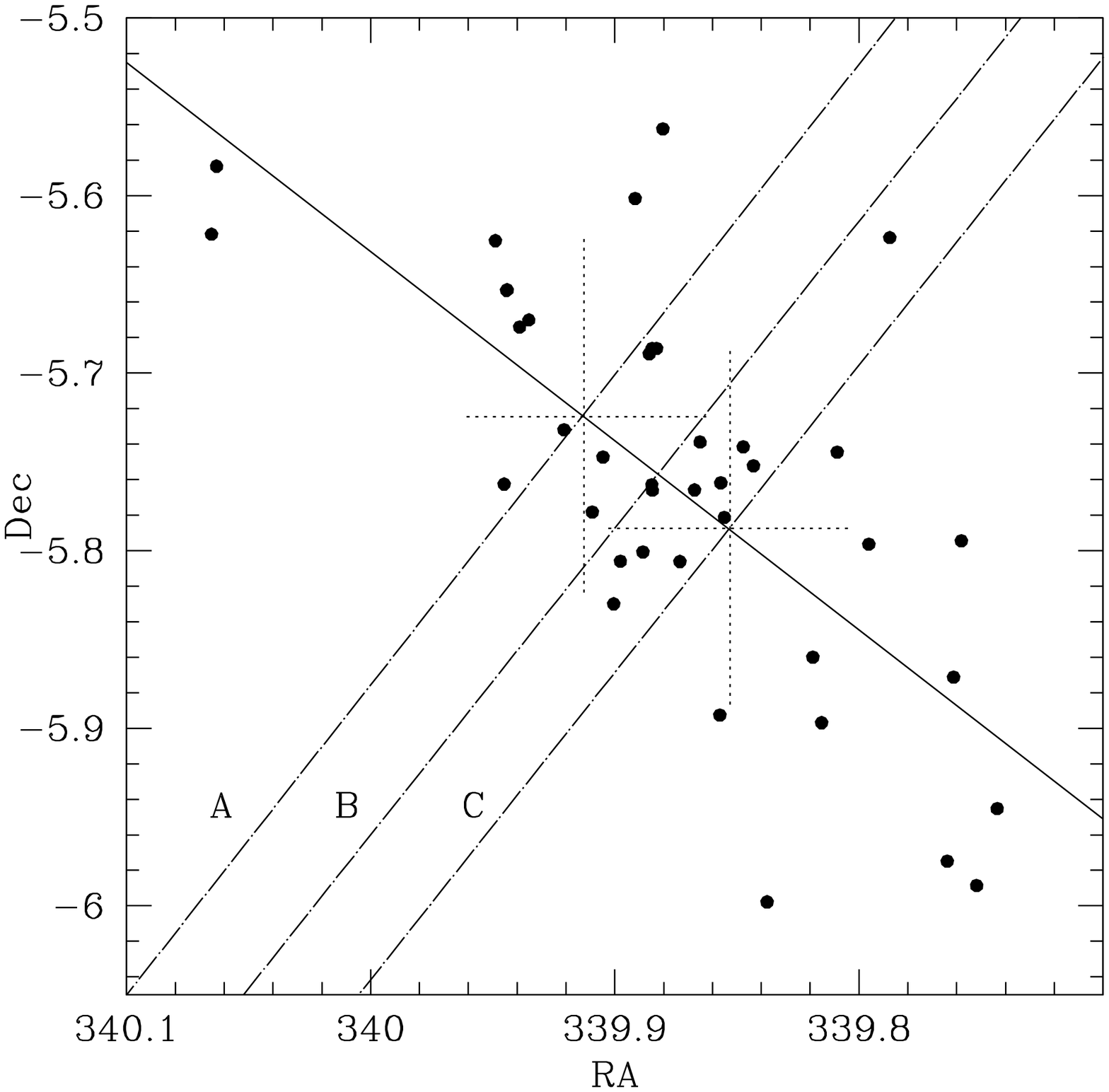}
\caption{ (left) The positions of emission line (solid dots) 
galaxies with detected H$\alpha$ emission
and non-emission line (crosses) galaxies observed in Abell 2465
with $70000 \leq cz \leq 76000$ \kms. (right) Positions of emission
line galaxies only, enlarged to show the central portion of the double
cluster. The XMM positions of the two X-ray clumps are marked with
the crosses. 
}
\label{emgalpos}
\end{figure*}

The emission line galaxies prefer to sit between the two 
subclusters with more near the SW clump. In 
Figure~\ref{emgalpos}, the left panel shows the positions of all
spectroscopically verified cluster members. The right panel plots only 
emission line objects and is an enlargement of the centre. The X-ray centres 
from Table~\ref{tab1objects} are indicated as crosses. The line passing
through the centres approximates the cluster's axis, and the lines
labled A, B, and C define stripes running perpendicular to this axis. The
line B marks the distance half way between the two subcluster centres. Using
this centre line, of the 39 galaxies in the figure, 24 are on the
SW side and 15 are on the NE side. However, the asymmetry is stronger in  the 
central region of the cluster. In strip BC, 13
emission galaxies lie close to the SW centre while only five are near the NE 
clump in strip AB. For non-emission line galaxies, the corresponding
numbers are 20 and 15 galaxies respectively.

This impression of asymmetry between the emission and non-emission line 
galaxy distributions 
was tested withthe two dimensional Kolmogoroff-Smirnov
test (Fasano \& Franceschini 1987; Lopes, Reid, \& Hobson 2007). Using the
{\sc ks2d2s} programme (Press \etal 1992) for the two-sample test
to compare the emission and non-emission samples
indicates weakly that the two types of galaxies are slightly different
at the 78\% significance level.    

There are more emission line galaxies near the centre of Abell 2465 than
expected for single galaxy clusters.
Balogh \etal (2004) and Rines \etal (2005) find an inverse
correlation between the number of emission line galaxies and density in
galaxy clusters. For their composite cluster, Rines \etal (2005; figure 2) 
find that the fraction of galaxies showing emission lines grows from 0
at the centre to 0.12 at $R_p = 0.5R_{200}$ with a mean near 0.06.
For Abell 2465, the numbers of galaxies with observed spectra 
within the circles in
Figure~\ref{fan.eps} which are projected radii of
$R_p = 0.53R_{200}$ (0.63 Mpc) from the
subcluster centres, are 27 and 22 for SW amd NE respectively. The
corresponding numbers of emission line galaxies are 9 and 4 giving 
fractions of emission line galaxies within these circles
equal to 0.33 and 0.18. Even allowing for infall
interlopers, these fractions appear different for Abell 2465 and the 
other clusters.

One question is whether the high fraction of star-forming galaxies could be due
to a bias in selecting more emission line galaxies because it is easier to get
their redshifts. With the current data, this can be answered only roughly
given the broad selection criteria and the employment of different 
spectroscopic instruments. A lower limit to the emission fraction can be
estimated using the numbers in \S 2.2. Firstly, 199/359 = 0.55 of the
galaxies with redshifts should be cluster members. Secondly, if all 93
failed redshift targets are non-emission galaxies, 55\% or 51 would be
cluster members. Thirdly, adding this to the number of observed cluster 
galaxies gives the lower limit to the fraction of star-forming cluster
members to be 55/(149+51), or 28\%. This is still in excess of the fraction,
with mean near 6\% for single clusters discussed above.

\section{THE LUMINOSITY FUNCTIONS OF THE CLUMPS}
In this paper, the statistical method of measuring all the galaxy photometry
in the field and then subtracting the contribution of the background is
employed using the nearly one square degree CFHT images.
Photometry of the cluster is confined to the inner $22 \farcm 4 \times
18 \farcm 0 $ portion
of the image and the sky background is estimated from the outer
part of the images. The data provided by the programme {\sc Sextractor} 
(Bertin \& Arnouts 1996; Bertin 2009; Holwerda 2005) described above
were employed for separating stars and galaxies. The red sequence of
the cluster described in $\S 2.1.3$ was used to find the cluster members.

\subsection{Background Galaxy Determination}
\begin{figure} 
\includegraphics[width=0.5\textwidth]{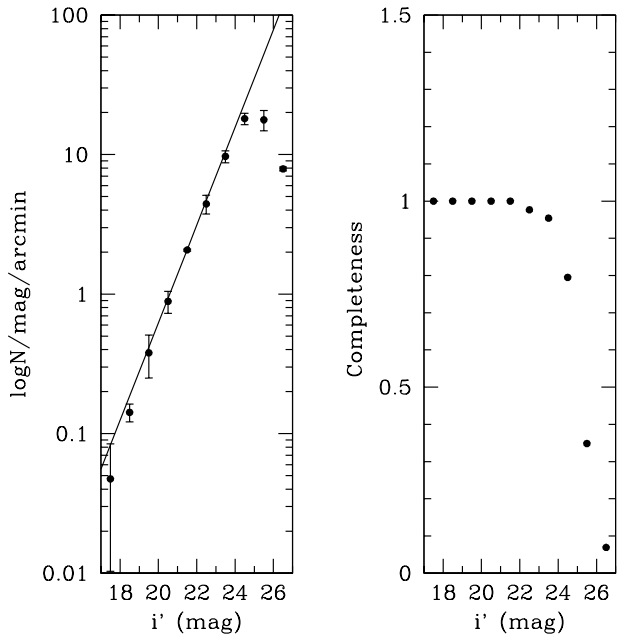}
\caption{ (left) Comparison of $i'$ band number counts found
in this study with McCracken's \etal (2001) deeper
counts shown as the line. (right) Adopted $i'$ completeness function for the 
galaxy number counts.
}
\label{backplot}
\end{figure}

The background, its errors, and the incompleteness 
were estimated in two steps. First
the resulting catalogue of galaxies was binned in magnitude using no
colour cut. The
outer portion of the CFHT $i'$ image was binned in one magnitude
intervals running from $i' = 16.0-26.0$. The total area used was 
3069 arcmin$^2$. The number of detected objects and their colours 
was 89579. The resulting number counts of the background along with their
$\sqrt{N}$ errors are shown in Figure~\ref{backplot},
where they are  compared to the number counts given
by McCracken \etal (2001) from deeper CFHT $i'$ imaging, and
Wilson (2003) who used Cousins $I$ 
band data from the same telescope. 

These data agree well for
$19 \la ~i' \la 23$. For $i'$ brighter than 19, small number statistics 
dominate and it is assumed that the numbers of detections is complete.
For $i'$ fainter than 23, the completeness is estimated by taking the ratio
of the present measurements and the McCracken \etal (2001) relation. The
resulting estimate of completeness is also shown in 
Figure~\ref{backplot} and is
1.0 for $i' =  17-22$. For fainter sources, it drops and
this relation was adopted for the completeness.

The second step was to apply the same colour cut found for the red sequence
to the background data. In addition to counting errors, there are the
effects of cosmic variance in the background. To estimate these, the
background was divided into four subsections of average 771 arcmin$^2$.
The variance of these measurements was used to estimate the error in the
background.

\subsection{Luminosity Functions of the Two Cluster Centres}
\begin{figure} 
\includegraphics[width=0.5\textwidth]{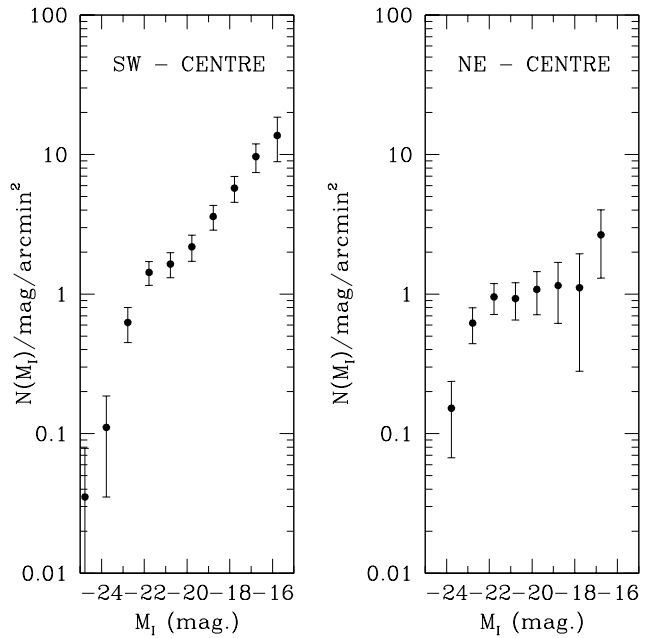}
\caption{  Luminosity functions for the central regions 
($R_p < 0.63$ Mpc or 2.75$\arcmin$) of the SW and NE clumps.
}
\label{HISTO_C}
\end{figure}

The luminosity functions (LF) of the SW 
and NE clumps of Abell 2465 differ. They were
obtained for the circular regions within $2 \farcm 75$ of the cluster
centres as defined in Table~\ref{tab1objects}. 
Identical colour cuts and the same background
subtraction were applied to the data of both clusters. 
The magnitudes $M_I$ are converted to
luminosity using $L_I = 10^{-0.4(M_I - I_{\odot})}$ where $I_{\odot} = 4.08$
(Binney \& Merrifield 1998) and $M_I= i' +DM-A_I-K_I(z)$,
using the data in Table 1. 
Bins fainter than $i' = 25$ were rejected due to the large incompleteness
correction. The results shown in Figure~\ref{HISTO_C}, where the error bars
include both background and counting errors, show a significant 
difference between
the two clumps, although both are within the range of luminosity
functions found for different clusters.

The bright portions of the two LFs are similar, but the SW clump
has a substantially larger number of galaxies fainter than $M_I \approx -21$.
This also can be seen by visual inspection of the central regions of the
two clumps.

\begin{figure} 
\includegraphics[width=0.45\textwidth]{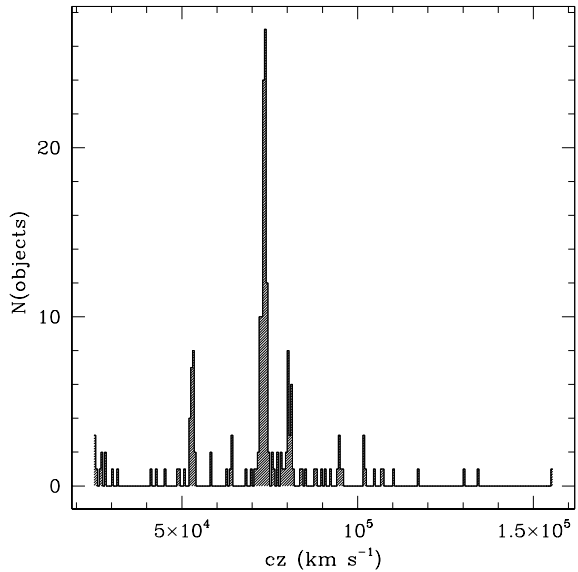}
\caption{  Redshifts of all galaxies centred on the SW clump.
The strong peak near $z = 0.245$ comes from the Abell 2465 galaxies.
}
\label{swallv}
\end{figure}

Although this statistical method provides a relative measure of the
LFs of the two clumps, it is less secure than using redshift based LFs.
A possible explanation of the differing LFs of the two clumps might be
that a distant cluster lies
behind the SW clump. Figure~\ref{swallv} shows the
distribution of redshifts centered on the SW peak and although two weak
peaks occur near $cz = 53000$ and 80000 \kms there is no
significant structure for $cz < 120000$ \kms.

\begin{table}
\caption{Relative Normalization factors for the luminosity functions and
assumed $(B-I)$ colours}
\label{lumfunc.tab}
\begin{center}
\leavevmode
\begin{tabular}{lllll}
\hline
      & E &S0 &Sp &dE \\
\hline
$(B-I)$&3.16&2.79 &2.16& 2.5\\
SW clump &0.5&2.5&1.0&4\\
NE clump &0.35&1.75&0.7&0.2\\
\hline
\end{tabular}
\end{center}
\end{table}

Were a substantial number of redhifts available for $i'$ fainter than
20, this possibility could possibly be sorted out. 

\begin{figure*} 
\centering
 \includegraphics[width=0.45\textwidth]{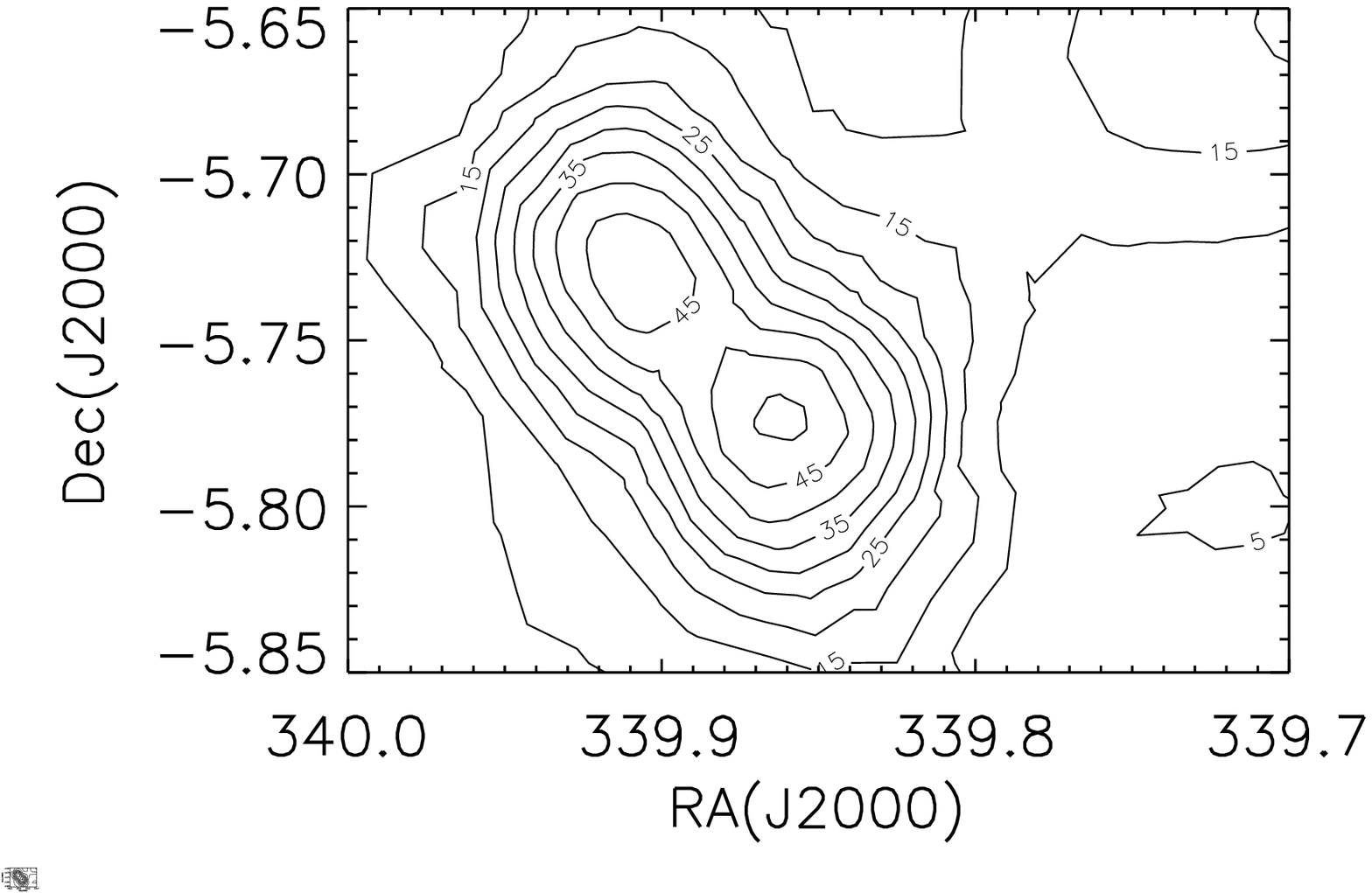}
 \includegraphics[width=0.45\textwidth]{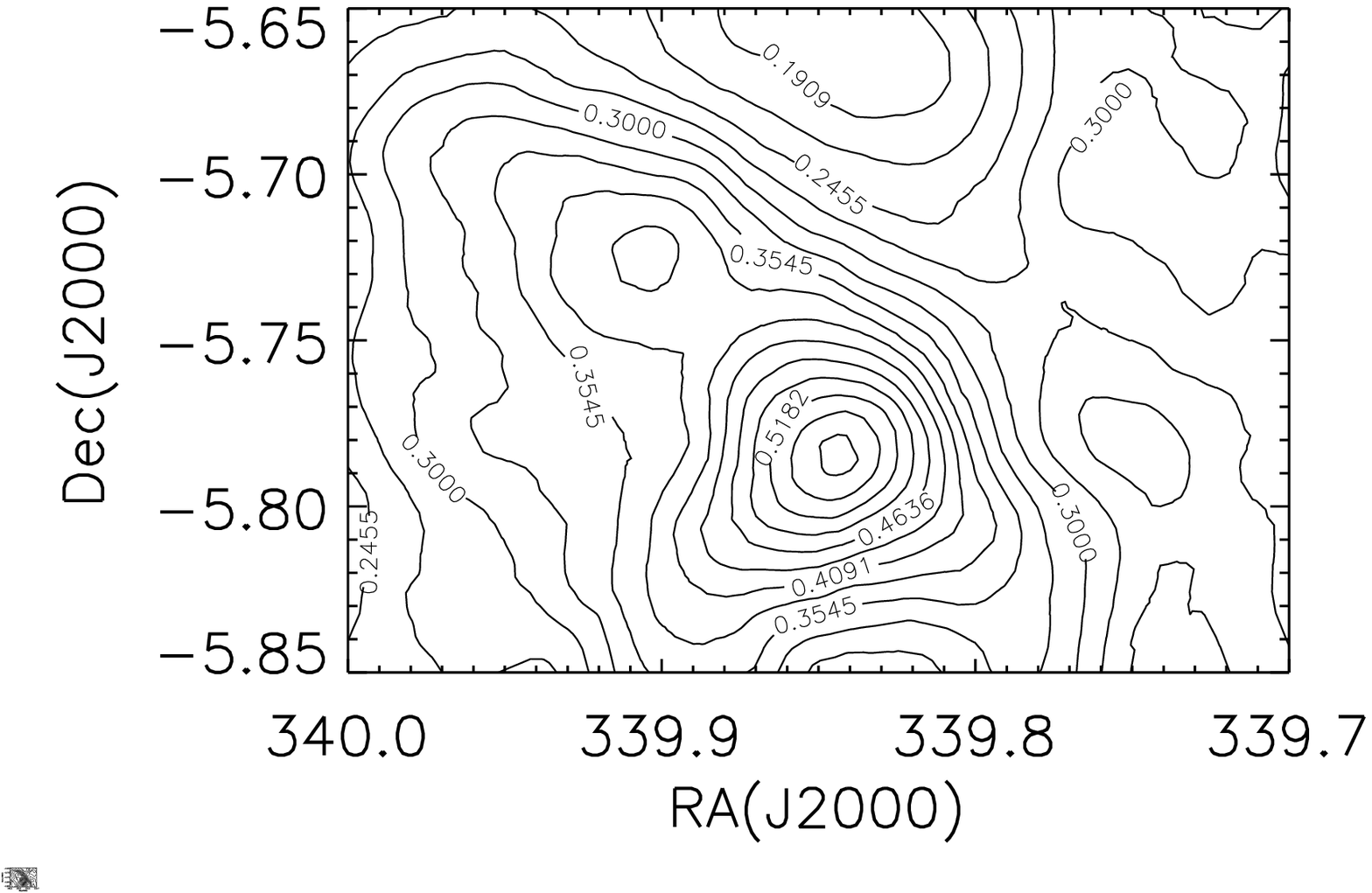}
\caption{  $I$-band isophotes for galaxies in the field of
Abell 2465 constructed with the adaptive kernal method. (Left) All galaxies
on the red sequence shown in Figure~\ref{redseq.eps} brighter 
than $M_I = -20.0.$ 
(Right) All the fainter galaxies with $-16.0 > M_I \ge -20.0$. The 
dominance of the SW clump  compared with the NE is notable. The scale of
the isophotes is in arbitrary units.
}
\label{contours}
\end{figure*}

Nevertheless, a distant cluster might show detectable differences in centre
and in shape compared to the foreground cluster. 
To look for these effects, the galaxy sample was divided
into bright ($M_I < -20.0$), and faint ($-16.0 > M_I \ge -20.0$) galaxies.
Their values of $L_I$ were converted from $i'$ as described above
and isophotes were constructed using the Silverman (1986) adaptive kernal
smoothing method. Figure~\ref{contours} 
has bright galaxies on the left. Both clumps show roughly 
circular isophotes and the
peak of the SW clump is the higher. The isophotes for the faint 
galaxies on the right differ. Here the
NE clump is diminished relative to the SW clump, 
which is stronger. Using the faint galaxies, the peak of the SW
clump is only slightly farther away from that of the NE clump along the axis
joining them by $\approx 0\fdg02$ or 0.3 Mpc.

The red sequence provides a second test for a 
distant cluster behind the SW clump.
Such a cluster might have a typical $I$-band Schechter 
function ($\alpha = -1.27, M^* = -21.66$; \eg Harsano \& De Propis 2009).
Assuming that the SW clump's LF is the same as that of NE plus the
Schechter function, the distance modulus of the distant cluster should be
larger by $\Delta DM \approx 3.2$ mag., or its distance modulus is
43.6, and  $z \approx 0.85$. Although
such a cluster lies beyond the currently available 
spectroscopy, at this $z$ the change in $K$-corrections alone
would make the distant cluster's $(r'-i')$ colour redder by about 1.3 mag.
(\eg Fukugita \etal 1995). This effect is well illustrated by Gilbank \etal 
(2008) for composite clusters in $(R-z')$ for $z$ = 0.4 to 0.9 which
include evolutionary effects. This shifting of the red sequence
could distort the shape of the red sequence of the 
SW region relative to the NE. One complication is that at increasing $z$,
the blue sequence grows stronger and by $z = 0.85$ it lies near the low
redshift red sequence. However, the red sequence remains the stronger and
should still be detectable. 

\begin{figure} 
\centering
\includegraphics[width=0.45\textwidth]{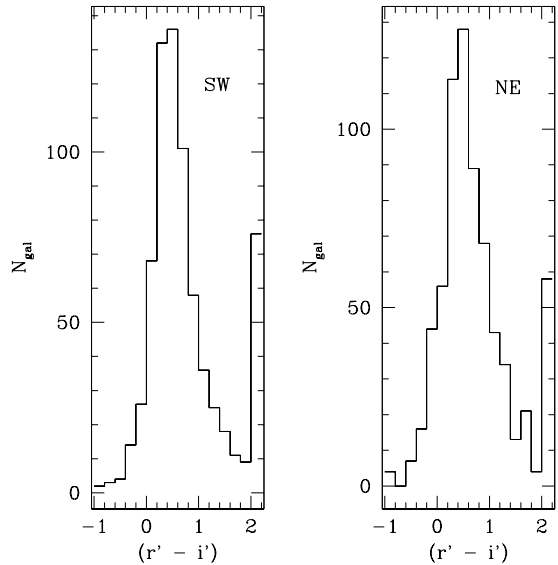}
\caption{  Histograms of the colour distributions for galaxies 
within $2 \farcm 75$ in
the two clumps of Abell 2465 with apparent magnitudes $20 .0 \le i' \le 24.4$.
}
\label{2colours_histo}
\end{figure}

To look for the shifted red sequence of
a distant cluster, galaxies with $20 .0 \le i' \le 24.4$ were
binned in $(r'-i')$ for both clumps within $2 \farcm 75$ of their centres.
As shown in Figure~\ref{2colours_histo}, there is no apparent enhancement 
in $(r'-i')$ near 1.7 for the SW region compared to
the NE. Applying the two-sample Kolmogorov-Smirnov test (Press \etal 1992)
supports the null hypothesis that the two samples are the same at the
99.8\% level so a substantial cluster directly behind the SW clump
seems unlikely. In the future, bluer colours would provide a more sensitive
test.

\begin{figure}
\centering
\includegraphics[width=0.5\textwidth]{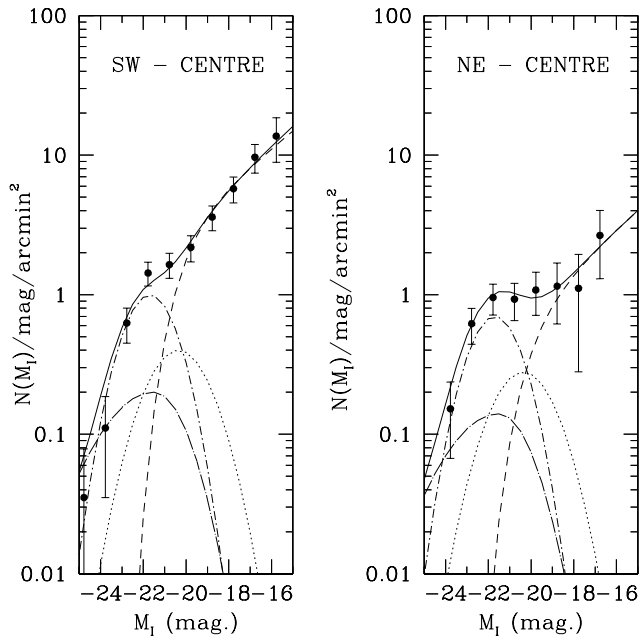}
\caption{  Fits to the luminosity functions in Figure~\ref{HISTO_C}
 using
the prescription given in Table~\ref{lumfunc.tab}. 
long dash - dot ellipticals, short dash - dot S0s, dots spirals,
and short dashes dEs. The solid curve is the sum of the individual components.
}
\label{lumfunc}
\end{figure}

In a test to fit different luminosity functions to
each clump, two approaches were employed. The first uses
the prescription of Bingelli, Sandage, \& Tammann (1988), also similar to
Wilson \etal (1997). The E, S0, and spiral constituents are fit by
Gaussian functions and the dEs and dIrr follow Schechter functions.
Here the fitting values of Jerjen \& Tammann (1997) are used
for the Gaussians and Bingelli \etal (1988) for
the Schechter functions. The $B$-band magnitude zero-points were adjusted
using mean $(B-I)$ colours for each galaxy type from Fukugita
\etal (1995) and Smail \etal (1998) for the dEs.
In Figure~\ref{lumfunc} reasonable fits to the observed luminosity
functions of the two clumps are achieved. The relative proportions of 
the galaxy types given in Table~\ref{lumfunc.tab} are 
similar for the two clumps, whilst the relative number of dEs is
approximately five times higher in the SW clump compared to the NE. It 
should be noted
that $\alpha = -1.35$ for the dE Schechter function is employed rather than
the steeper $\alpha \approx -1.6$ to -2 predicted by $\Lambda$CDM.

The second approach uses the double Schecter function
Popesso \etal (2006) and others found that fits many clusters. 
Using published parameter values, these LFs do not agree well with those 
of Abell 2465. Other authors, \eg
Wolf \etal (2003) and Christlein \etal (2009) also find that simple
luminosity functions do not fit available data and obtain luminosity
functions similar in shape found here. They interpret the 
luminosity functions as the sum of early types plus a 
rising late-type component composed of mostly faint blue star-forming 
galaxies.

Independantly of the fitting method, the outstanding feature is
the difference in the numbers of faint galaxies following the dE
Schechter function. There is an excess in the SW
clump which is the more luminous and slightly less massive, while the NE
clump has fewer faint galaxies. Additional spectroscopic data are needed to
establish the nature of the faint galaxies in the SW clump.

\section{LIGHT PROFILES OF THE CLUMPS}

The luminosity functions data are also employed
to construct the light profiles. 
The 
apparent $i'$ magnitudes were converted to
absolute $M_I$ Cousins magnitudes using the transformation
in \S 4.2. The total light, $L_p$, within circular apertures of radius
$R_p$ was
measured
starting from the cluster centres increasing the radius in $0 \farcm 15$ 
steps for projected radii. For $R_p \leq 2 \farcm 75$ circular areas were 
taken but for larger radii, the areas were corrected for the 
area of the missing segment that overlaps the other clump.  
The sums inside the aperture radii,
or growth curves give smoother curves than the projected luminosity. 

\begin{figure*}
\centering
 \includegraphics[width=0.45\textwidth]{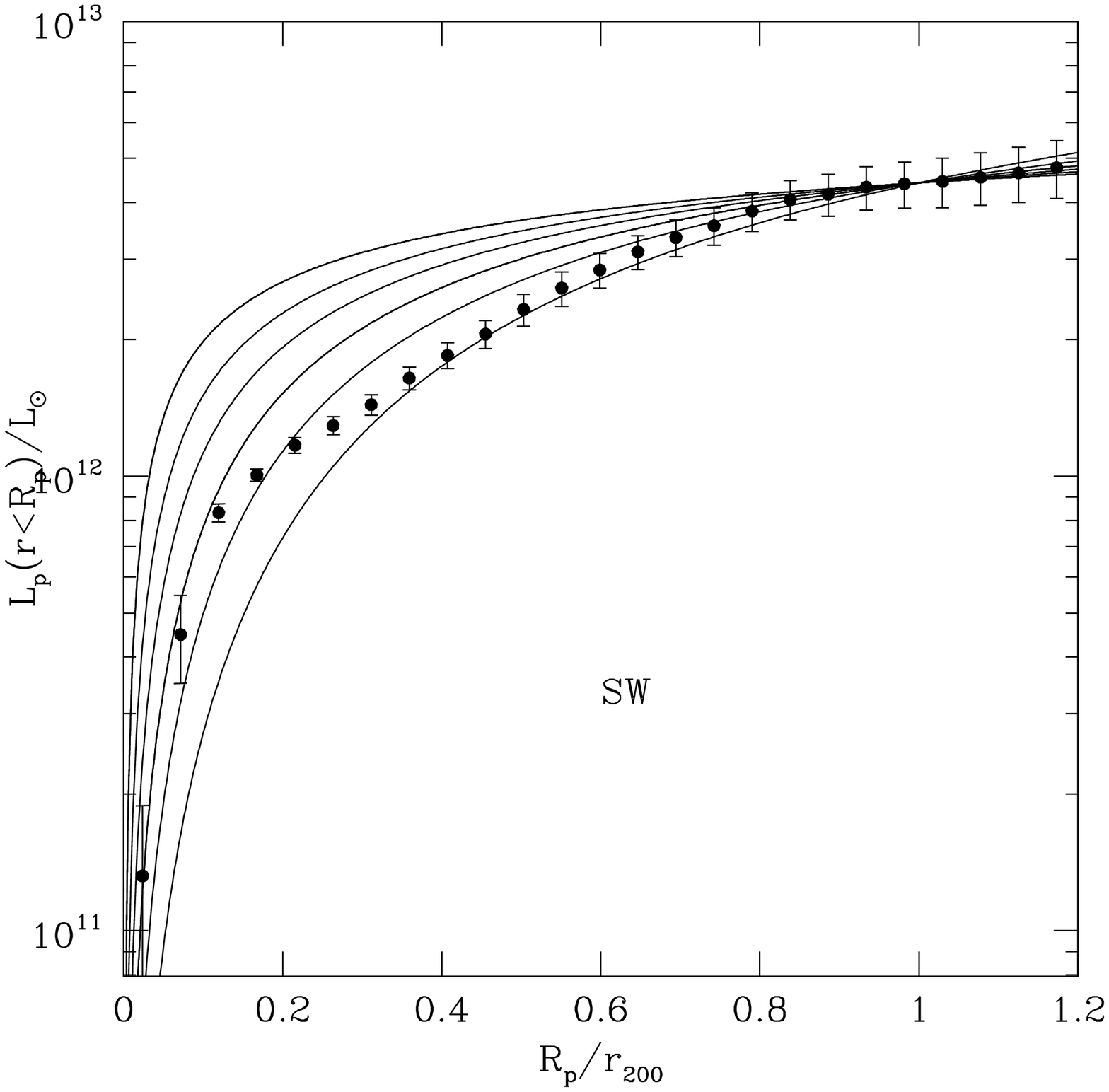}
 \includegraphics[width=0.45\textwidth]{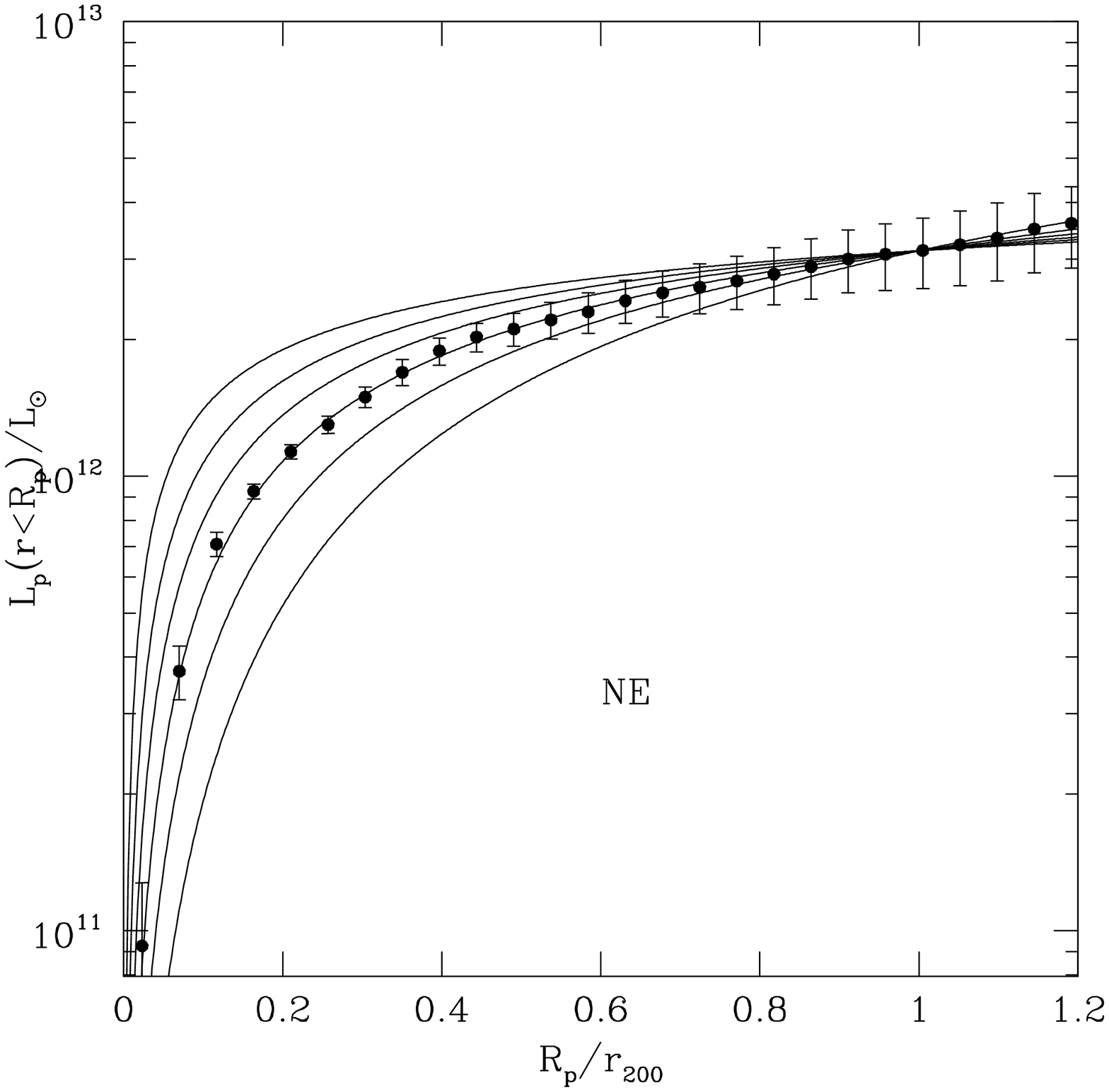}

\caption{  The total projected light inside radius $R_p$ 
for the two 
clumps of Abell 2465 and their errors. Curves for NFW profiles 
with different concentration indices are
superimposed; from top to bottom: $c =$ 100, 40, 20, 10, 5, and 2. SW has
$L_p$ = 4.4$\times10^{12} L_\odot$ at 
$r_{200} = 1.21$ Mpc and fits with
$c \approx 4$, while NE has $L_p$ = 3.8$\times10^{12} L_\odot$ at 
$r_{200} = 1.25$ Mpc with $c \approx 10$.
}
\label{NFWprofiles}
\end{figure*}

The choice of the cluster centre affects results near the origin,
but its influence decreases with radius. The BCGs and the X-ray centres 
of the NE clump differ by $\sim$22$\arcsec$. Therefore the aperture profiles 
were constructed several times taking different centres
within $\pm$22$\arcsec$ and averaged. This
affects the central few points, but becomes negligable for
radii larger than $\sim 0 \farcm 45$, where the background's uncertainty
dominates. The errors are estimated from the variance of the
different counts and the counting statistics. The resulting curves are shown in
Figure~\ref{NFWprofiles} where they are compared with NFW growth curves 
for different concentration indices.

Several profile functions are known to fit galaxy
clusters (\eg Katgert \etal 2004). Althought the purpose here is not to
determine the optimal profile function, it is convenient to compare
NFW profiles (Navarro, Frenk \& White 1997)
to the data in Figure~\ref{NFWprofiles}. The curves,
assuming that $M/L \approx \rm{const}.$, are for different
concentration parameters, $c$, computed using the formula for the projected
mass, 
\begin{equation}
M_p(R) = g(c)M_v\bigg[\frac{C^{-1}[1/(c\tilde{R})]}{|c^2\tilde{R}^2-1|^{1/2}}] 
+ \ln\big(\frac{c\tilde{R})}{2}\big)\bigg],
\end{equation}
({\L}okas \& Mamon 2001), where $\tilde{R} = R/r_v$, $M_v$ is the
virial mass, 
\begin{equation}
g(c) = \frac{1}{\ln(1+c)-c/(1+c)},
\end{equation}
and $C^{-1}(x) = \cos^{-1}(x)$ for $R > r_s$ and $\cosh^{-1}(x)$ if
$R < r_s$. 
The best fitting NFW profile was found using the Kolmogoroff-Smirnov test
(Kreyszig 1991; Press \etal 1992) on the projected growth curve in
Figure~\ref{NFWprofiles}. Note that data are displayed in a
semi-log plot whilst the KS test is done linearly.  In
these determinations, the spectroscopic values of $r_{200}$ were used.
SW follows curves with $c = 4 \pm 2$ and NE lies closer to $c = 10 \pm 5$.
However, although the single NFW profile fits the NE region relatively 
well, agreement is poorer for the SW clump,
particularly near $R_p/r_{200} \approx 0.3$ where the data rise 
compared to the nearby smooth NFW curves for different values of $c$.

\begin{figure} 
\centering
\includegraphics[width=0.45\textwidth]{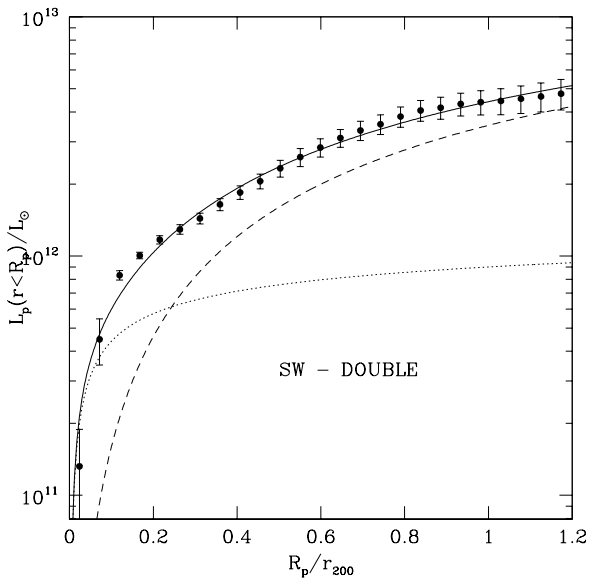}
\caption{  Fit to the growth curve of the SW clump of Abell 2465
data given in Figure~\ref{NFWprofiles} using the double model discussed in
the text. The dashed curve is for a NFW profile with $c$ = 1.0 and the
dotted curve is for $c$ = 120. The solid curve is their sum.
}
\label{doubleprof.eps}
\end{figure}

Viewing Figure~\ref{cluster.eps}, this rise should result from the
bright BCG complex near the centre of the SW clump. Consequently, a more
complicated composite profile containing a bright core 
yields a better fit. However, which
combination of profiles is the best is difficult to decide with the present
data. One possibility shown in Figure~\ref{doubleprof.eps} is to add
two NFW profiles, one with $c$ = 120 which gives a sharp core and the
other with $c$ = 1.0 which produces an extended outer region. This
confirms that the SW clump is more centrally concentrated as it also
is in X-rays.

Measurements of $c$ based on mass give somewhat different values for $c$.
Katgert \etal (2004) find $c=4^{+2.7}_{-1.5}$ from a cluster
ensemble, while Biviano \& Girardi (2003) find $c\approx 5.6$ and Carlberg 
\etal (1996) obtain $c\approx 4$. Thus $c$ for the SW component
might  lie within the normal range except for its core,
but for NE clump $c$ is larger. Theoretical
values of $c$ based on the buildup of dark haloes in CDM models (\eg
Zhao \etal 2003) agree with the lower value of $c$, so the higher value for
NE is difficult to explain, although the value of $c$ depends on the
accretion rate of the cluster. 

The value of $c$ does depend on the choice of $r_{200}$. However, to achieve
a fit resulting in $c = 5$ for the NE clump
requires lowering $r_{200}$ to half its
spectroscopic value, or $6 \sigma$. 
If the mass distributions in the subclusters
of Abell 2465 differ from those found here for light, this may imply
that they may have been disturbed by the merging process.

As seen in Figure~\ref{NFWprofiles}, the total 
luminosities are $L_I = 4.4 \pm 0.6 \times 10^{12} L_\odot$ for SW
and $L_I = 3.8 \pm 0.7 \times 10^{12} L_\odot$ for NE. Using the mean of the 
virial and X-ray masses, gives mass-to-light ratios in the $I$-band of
$M/L = 84\pm 12$ and $M/L = 112\pm 20$ respectively. 
Noting that $(R-I) \approx 0.7$ for early-type galaxies, in the $R$-band
(Fukugita \etal 1995), $\Upsilon_R \approx 1.9\Upsilon_I$ so
the mass ratios lie marginally within the range for single galaxy clusters 
which is 
$\Upsilon_R$ = 200$\pm$ 50 in the $R$-band (Binney \& Tremaine 2008). 
The more massive NE clump has the higher $M/L$ while the SW clump has the more
concentrated core.

\section{DISCUSSION}
The mass ratio near unity indicates that the components of Abell 2465
are undergoing a major merger. However,
the question that needs to be answered is whether Abell 2465 is beginning
the merger or if a collision has already occured.
The main properties of this double galaxy cluster have both some 
normal and some unusual properties. On one hand, the masses and virial
radii are similar to single clusters as are their velocity dispersion 
anisotropies. On the other hand, the differences in
their luminosity functions, $M/L$, and the presence of many star-forming 
galaxies located between the two clumps offer additional clues to the
processes involved in their earlier interaction which tend to 
favour the past collision hypothesis. 

\subsection{Separation of the Baryonic and Collisionless Components}
After pericentric passage in galaxy cluster collisions,
a generic result is that 
the highly heated baryonic gas is temporarily retarded relative to
the collisionless dark matter and galaxies with the result that
the X-ray centres are closer together than the dark matter centres. The
separation is expected to be
smaller for nearly equal mass clusters than it is for higher mass
ratios (\eg Poole \etal 2006; Tormen \etal 
2004). The gas cools and re-merges with the collisionless
components at later times. These effects are 
observed in several recent
($\tau_{coll} \la 0.1 - 0.3$ Gyr) collisions with higher mass ratios, notably 
1E0657-56 (Clowe \etal 2006), Abell 2146 (Russell
\etal 2010), and MACS J0025.4-1222 (Brada\v{c} \etal 2008b),
where X-ray emission is 
between the dark matter clumps as revealed by lensing and the 
galaxies. 

For Abell 2465, the X-ray and BCG centres are in Table~\ref{tab1objects}. 
Displacements occur along the axis in Figure~\ref{emgalpos} 
joining the two clumps,
putting the X-ray peaks between the BCGs. These amount to $22 \farcs 1$
(85 kpc) for the NE and $2 \farcs 4$ (9kpc) for the SW. The isophotal peaks 
from the galaxies in Figure~\ref{contours} are more unreliable. These change by
$\sim \pm 25 \arcsec$ depending on whether different magnitude ranges or
galaxy numbers are used to construct the contours. These place the centres
of the galaxy distributions near or slightly inside the X-ray peaks along 
the axis. Relative to the $5 \farcm 5$ separating the clumps, displacements
are small compared to the colliding clusters above. Thus, while the 
separation of the components is consistent with a past collision, 
using the X-ray and BCG locations,
their small separations suggest that the NE and SW subclusters have
not recently collided.

\begin{figure}
\centering
\includegraphics[width=0.45\textwidth]{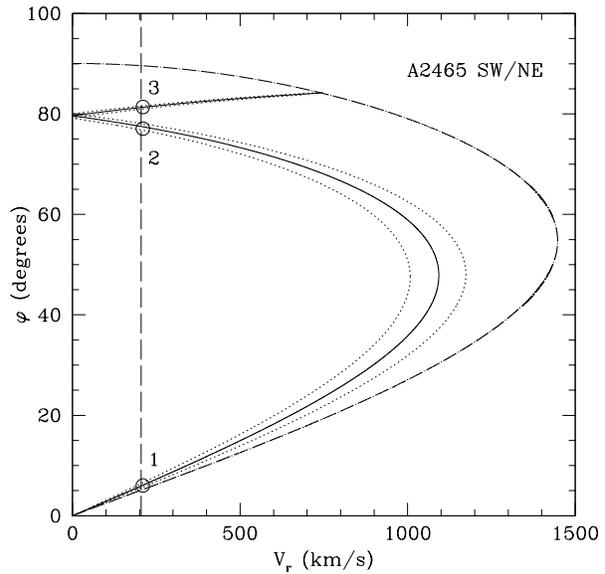}
\caption{  Radial infall model solutions for Abell 2465.
Rightmost (dot-dash) curve is limit of bound solutions; solutions to its right
are unbound. Solutions for $8\times10^{14} M_\odot$ 
(solid) and $9\times10^{14}M_\odot$ and $7\times10^{14}M_\odot$ (dotted)
lie on either side. The vertical dashed line shows the
measured value of $V_r$ and circles mark the three possible
solutions indicated by the numbers.
}
\label{radialinfall.eps}
\end{figure}

\subsection{Nature of the Interaction}
The radial infall model (Beers, Geller, \& Huchra 1982) 
has often been used as a first approximation to study head-on collisions,
despite its obvious limitations, including
neglect of dynamical friction and gas dynamics. It
is an analytic solution based on the Einstein-de Sitter
cosmological model and has been employed by many investigators 
(\eg Gregory \& Thompson 1984; Beers \etal 1991; Scodeggio \etal
1995; Colless \& Dunn 1996;
Mohr \& Wegner 1997; Donnelly \etal 2001; Yuan \etal 2005;
Hwang \& Lee 2009) and will only be outlined here. Briefly, 
two mass points with total mass, $M$, following radial
orbits are assumed and $R$, their separation, $V$,
the relative velocity, and the time, $t$ are given parametrically
in terms of the development
angle, $\eta$. The masses coincide at pericentre when
$\eta = 0, 2\pi,...$ and are at apocentre when
$\eta = \pi, 3\pi,...$ The solution requires $M$, the projected
distance, $R_p$, and $t_0$, the system's age, 
usually set equal to the age of the Universe. Since the inclination angle
$\varphi$, is unknown, the system of equations can be written in terms of 
$R_p = R \cos \varphi$,
the projected distance betwen the masses, radial velocity difference,
$V_r = V \sin \varphi$ and $M$. The bound case for the
two masses obeys $V_r^2 \leq 2GM\sin^2 \varphi \cos \varphi$ and
three solutions are possible: two collapsing or ingoing and one 
expanding or outgoing. The $M$ and $V_r$ of \S3 place
the cluster pair
within the bound region and hence the unbound case is not considered.

Using
$M=8\pm 1 \times 10^{14}M_\odot$, $R_p$ = 1.265 Mpc, and 10.895 Gyr, 
the age of the universe at $z=0.245$, one can
construct curves of the solutions within
the permitted regions shown in Figure~\ref{radialinfall.eps}. 

With $V_r = 205$ \kms, the three possible solutions are: (1)
$\eta=5.17$ radians, $\varphi=5.95$, $R=1.31$ Mpc, $R_m=4.53$ Mpc, and 
$V= -1978$ \kms.
(2) $\eta=3.53$ radians, $\varphi=77.5$, $R=6.01$ Mpc, $R_m=6.07$ Mpc, and 
$V=-210$ \kms.
(3) $\eta=2.68$ radians, $\varphi=81.3$, $R= 8.59$ Mpc, $R_m=8.81$ Mpc, and 
$V=+208$ \kms.
Intuition suggests that either solution (2) or (3) might be preferable 
given the low observed
value for $V_r$, placing the two clumps near apocentre, but
further data are required.

Numerical simulations provide a better time scale estimate 
of galaxy clusters, including galaxies, baryonic and dark matter.
The colliding models typified by \eg Roettiger \etal (1996; 1997), 
Ritchie \& Thomas (2002), Poole \etal (2006; 2008), etc while idealized and
differing in details, predict similar effects in  the collisions 
and suggest shorter times between collisions than the radial infall
model predicts.

Published merger models focus on X-ray data and employ
several simplifications, including \eg assumed initial density profiles,
velocities, and spherical symmetry. Nevertheless, three effects 
common to the models are important
for this study. (1) Head-on and off-centre simulations
generally merge soon after their second
crossing, indicating that the two subclusters in Abell 2465
have collided at most once or are merging for the first time. (2) The
gas components' temperatures spike when the two centres cross and
cool rapidly. Following Ritchie 
\& Thomas (2002), the collision of $M=4\times10^{14}$M$_\odot$ equal masses,
close to that estimated for Abell 2465, the gas temperature 
reaches $\sim10$ keV and cools to $\sim3$ keV in about 3 to 
4 Gyr as the clumps separate to apocentre and begin reconverging.
The temperatures of \S 3.3, are consistent with this
time scale if the cluster is in the post-impact stage, but cannot be used to
prove the physical state of the merger.
(3) The collision produces an expanding impact disk of gas 
as two clusters pass through each other which breaks 
into two sections continuing outward with the main masses. Zu Hone, Lamb,
\& Ricker
(2009) have discussed ring formation. The dark matter components pass through
each other and the core regions are not disrupted.
 
Tormen \etal (2004) analysed
the evolution of merging galaxy cluster satellites from
hydrodynamical N-body simulations including dark matter and
baryonic gas. They fitted
the statistics of collisions including orbital properties,
velocity dispersions, gas temperatures, etc. as a function of the
satellite and main cluster pre-merger mass ratio, $m_v/M_v$. 
For $m_v/M_v \approx 0.9$, $\sim 2$ Gyr is required to reach apocentre. 
Consequently, this and the temperature aging indicate that the
time since the collision is of order $\tau \sim$ 2-4 Gyr.

The cluster light profiles and luminosity functions provide
additional clues to the interaction of the two components of Abell 2465. The
SW clump has a sharper core in both the visual and X-rays
as well as more faint galaxies shown in 
Figures~\ref{HISTO_C} and \ref{contours}.
The two well studied cluster collisions involving 
recent impacts, 1E0657-56 and Abell 2146 (Clowe \etal 2006; Russell
\etal 2010), have lower mass ratios of $m_v/M_v \approx 0.1$ and 0.3
respectively.
Both exhibit a relatively dense `bullet' emerging
from a more extended `target,' and
are near their pericentres, having collided
$\tau \sim 0.1 - 0.3$ Gyr ago, 
compared to
$\tau \sim 3$ Gyr for the cooler Abell 2465 now near apocentre. 
By analogy with these two objects, one
identifies the SW clump with its sharper optical and X-ray core as the 
bullet and the more extended NE clump as the target. The presence of the
1.4 GHz radio source and its higher mass further imply that the NE clump is 
the primary.

Although cluster cores are not disrupted, the outer regions can be modified.
The published simulations show complex behavior of the collisionless component.
Bekki (1999) and Roettiger, Burns, \& Loken (1993) 
demonstrate that the distribution of dark matter in the secondary
expands after the collision. Using the simple impact approximation
(Binney \& Tremaine 2008), energy changes accompanied by mass loss
of the impacting systems occur. Intuitively
one expects core regions to contract as they meet while the outer layers
of the clusters expand (\eg Aguilar \& White 1985; Funato \& Makino 1999).
Some of the off-centre models (\eg Poole \etal 2006; Ricker \& Sarazin 2001)
show that when the secondary core reaches apocentre and turns around, there
can be  a significant displacement of the cluster's outer region  which is
assisted by a gravitational slingshot similar to a trailing tidal trail.
Thus the SW clump might be expected to be surrounded by more debris than the
NE which accounts for its excess of faint galaxies.  

\subsection{On the star forming components}
Previous investigations have found inconclusive results for different
cluster pairs regarding 
star formation. Hwang \& Lee (2009) found enhanced star 
formation activity between the subclusters of Abell 168 which they concluded 
had passed through each other and none in Abell 1750 which may be in an
early stage of merging. Rawle \etal (2010) have found significant 
star formation in the bullet cluster using Herschel.
Caldwell \& Rose (1997) reported enhanced star forming in interacting clusters
as did Cortese \etal (2004) for Abell 1367, Ferrari \etal (2005) for Abell 
3921, and Johnston-Hollitt \etal (2008) for Abell 3125/28, but Tomita \etal
(1996) and De Propis \etal (2004) found no correlation for blue galaxies and
mergers. 
Ma \etal (2010) have described MACSJ0025.4-1225, which has several features
in common with Abell 2465, as a post-major-merger. They find enhanced 
numbers of starforming galaxies which they interpret to have been produced by
the merger. 
The emission line galaxies in Abell 2465, however, lends weight
to the hypothesis that cluster mergers enhance star formation. 

It has always been difficult to pin down the 
process or combination of processes leading to star formation in
galaxies. Bekki (1999) found that the time-dependent tidal gravitational field 
is an important effect that can trigger starburst galaxies in mergers.
Martig \& Bournaud (2008) also find that tidal fields in merging dense
cosmological structures and the outskirts of galaxy clusters can induce star
formation.
The presence of apparently distorted star forming galaxies with no detected
companions in the clusters could be consistent with this mechanism.

\section{CONCLUSIONS}
Spectroscopic and photometric observations of the double galaxy cluster
Abell 2465 are presented. There are five main  conclusions that can be drawn.

1. Concerning the cluster dynamics,
the virial masses of the two subclusters are
found from fuzzy clustering, which is used to estimate the probability of 
a galaxy's membership in each clump, with the result that 
$M_v = 4.0 \pm 0.8 \times 10^{14} {\rm M}_\odot$ for the NE member and
$M_v = 3.8 \pm 0.8 \times 10^{14} {\rm M}_\odot$ for the SW member and
the virial radii are $r_{200} = 1.21 \pm 0.11$ Mpc and 
$1.24 \pm 0.09$ Mpc for NE and SW respectively. The masses compare
well with those from X-ray scaling relations that also give temperatures
of 4.1 $\pm 0.3$ and 3.75 $\pm 0.2$ keV respectively. The velocity 
difference between 
the two subclusters is found to be $\Delta{V}$ = 205 $\pm$ 149 \kms which 
confirms that they are related. Measurement of the clusters' 
velocity dispersions with radius assuming spherical symmetry indicate
that the anisotropy parameter, $\beta$, is low.

2. There is an excess of star forming galaxies showing emission lines.
Of cluster members observed spectroscopically
in Figure~\ref{diagnostic.eps}, 37\% have
detectable H$\alpha$ emission. These have the properties of
star forming galaxies. There are more emission line objects in
the SW clump than in the NE clump and there appears to be
more emission line galaxies than non-emission between the two clumps.
This does not seem to be explained by a selection bias.
There is no evidence for strong AGN activity in Abell 2465. This
number of emission line objects between the clump centres is unusual when
compared to single galaxy clusters. 

3. The $r'$ and $i'$ magnitudes show well defined red sequences in
each subcluster. The luminosity functions determined within the central 0.6
Mpc of each clump indicate a normal mixture of galactic types. However,
the SW region has more galaxies fainter than $M_I = -20.0$
than its NE companion. This could result from their collision
or otherwise would suggest different formation histories.
The possibility of a background cluster needs to be further checked.

4. The light profiles of both components measured as growth curves were
fitted using NFW profiles. The NE clump is fit with a 
somewhat high concentration parameter $c = 10$, although this depends
on the adopted virial radius. The SW clump is
fit rather badly with $c = \sim 4$ and needs a profile with a more
compact core. A better fit is a sharp core ($c$ = 120) surrounded by an 
extended outer region ($c$ = 1.0). This is consistent with 
Figure~\ref{XMMNVSS} and published ROSAT data showing that the X-ray
core radii differ with $r_c$ of NE being about three times larger than
that of the SW and indicates that SW has a cooling core. The derived 
$I$-band mass to light ratios are 
$\Upsilon_I = 84 \pm 12$ and $112 \pm 20$ which puts them in the normal
range for galaxy clusters.

5. A consistent picture of the collision of the Abell 2465 components is 
discussed. It is possible that the pair collided 2-4 Gyr ago and 
are now near maximum separation. The small displacements of the dark matter and
baryonic matter as judged by the X-ray data and distribution of the
galaxies is consistent with their re-merging after the collision, 
The high percentage of emission line galaxies
in the spectroscopic sample may be a consequence of the collision and
are the strongest argument for a past interaction, but this
might also be the case if the merger is just starting and interaction occurs
along the interface between the two clusters. More models that
include the dynamics of the galaxies would be helpful.

A weak lensing study of the two components of Abell 2465 is under way.

\section*{Acknowledgements}
Many of the galaxy spectra were obtained through the 
Service Observing Programme of Anglo-Australian Observatory and I wish to 
express my gratitude.
The imaging observations used in this paper were 
based on observations obtained with MegaPrime/MegaCam, a joint project of CFHT 
and CEA/DAPNIA, at the Canada-France-Hawaii Telescope (CFHT) which is operated 
by the National Research Council (NRC) of Canada, the Institut National des 
Science de l'Univers of the Centre National de la Recherche Scientifique (CNRS)
of France, and the University of Hawaii. This work is based in part on data 
products produced at TERAPIX and the Canadian Astronomy Data Centre as part of 
the Canada-France-Hawaii Telescope Legacy Survey, a collaborative project of 
NRC and CNRS. I thank the Canadian TAC for granting the 
time and the QSO team for obtaining the imaging data.
This research has made use of the NASA/IPAC Extragalactic Database (NED) which 
is operated by the Jet Propulsion Laboratory, California Institute of 
Technology, under contract with the National Aeronautics and Space 
Administration. Many thanks to Dr. J. J. Mohr and Mr. Dane Owen for help on 
the MDM photometry, Mr. R. E. Johnson for discussions, and Dr. Heinz Andernach
for comments.

\label{lastpage}

\end{document}